\documentclass[12pt]{article}

\usepackage{setspace,graphicx,epstopdf,amsmath,amsfonts,amssymb,amsthm,versionPO}
\usepackage{marginnote,datetime,enumitem,subfigure,rotating,fancyvrb}
\usepackage{hyperref}
\usepackage{float}

\usepackage[longnamesfirst]{natbib}
\usepackage{jfe}    
\usepackage{lipsum}

\usepackage[utf8]{inputenc}
\usepackage{comment}
\usepackage{setspace} 
\usepackage{bibentry}

\usdate

\usepackage[titletoc,title]{appendix}
\usepackage{comment}

\newtheorem{corollary}{COROLLARY}
\newtheorem{proposition}{PROPOSITION}

\newtheorem{lemma}{LEMMA}
\newtheorem{remark}{Remark}
\newcommand{\argmax}{\mathop{\rm arg\,max}}

\theoremstyle{definition}

\newtheorem{defn}{Definition}

\newtheorem{assumption}{Assumption}[section]

\title{\Large \bf Buy It Now or Later, or Not\\
\large Loss Aversion in Advance Purchasing}

\author{Senran Lin\thanks{\footnotesize China Center For Behavioral Economics and Finance, Southwestern University of Finance and Economics, 555 Liutai Avenue, Wenjiang District, Chengdu, Sichuan, 611130, China. Email: \href{mailto:senranlin@outlook.com}{senranlin@outlook.com}. I am grateful to Martin Dufwenberg for his guidance and support. I also thank Soo Hong Chew, Shuya He, Gavin Kader, RC Lim, Joshua Lanier, Rachel Mannahan, and Jin Sohn for helpful comments, and audiences at China Center For Behavioral Economics and Finance, Antai College of Economics \& Management. Financial supports from the National Natural Science Foundation of China  (No. 72033006) are gratefully acknowledged. }}

\date{ }

\begin{document}

\setlist{noitemsep}  
\onehalfspacing   

\maketitle

\thispagestyle{empty}

\bigskip
\centerline{\bf ABSTRACT}

\begin{doublespace}  

\noindent 

This paper studies the advance-purchase game when a consumer has belief-based loss-averse preference, introducing a novel perspective by incorporating reference updating. It demonstrates that loss aversion increases the consumer's willingness to pre-purchase. 
Moreover, the paper endogenizes the seller's price-commitment behavior in the advance-purchase problem. The analysis reveals that the seller will commit to his spot price even with no obligation to do so, a behavior previously assumed in the literature.

\end{doublespace}



\clearpage

\section{Introduction}

Advance purchases exist widely in various markets, from sports and concert tickets to holiday accommodations, consulting services, and meals. Within these markets, a consumer has two opportunities to make a purchase at different points in time. She can either choose to pre-purchase in the \emph{advance-purchase stage}, or she can decide whether to buy in the \emph{spot-market stage}, which is close to the time of consumption.
The first stage involves deciding whether to buy early under uncertainty related to factors such as the weather, work schedules, or the consumer's state of mind. These uncertainties often get resolved by the time of the spot-market stage \citep{dana1998advance,shugan2000advance}. 
For instance, consider a sports fan who is uncertain about attending a game. First, she must decide whether to buy a ticket a month in advance, when she does not know what the weather will be like. If she chooses not to buy early, she can wait and learn the weather report, and then make a more informed decision about whether or not to buy the ticket closer to the game day.

Given the dynamic nature of the advance-purchase problem, the consumer's expectation play a central role in decision-making. At the same time, empirical evidence shows that loss aversion, which is inherently related to reference dependence, significantly influences consumer behavior \citep{heidhues2008competition,kalwani1992consumer,winer1986reference,hardie1993modeling}. Consequently, it is reasonable to consider a loss-averse consumer whose reference is shaped by her own expectations. This behavior pattern is characterized within the framework of the utility model by \cite{kHoszegi2006model,KR09}, hereafter referred to as the KR utility. In this context, an intriguing question arises: 
How does a consumer’s loss aversion influence her advance-purchase behavior and, hence, the seller’s optimal pricing policy?

An inherent property in the dynamics of advance-purchase decision-making is that the consumer's expectations evolve in response to new information. Hence, specifying how the consumer's reference is determined at each stage becomes essential.
Furthermore, some empirical evidence suggests that a player's reference points can update rapidly, effectively diminishing the impact of prior beliefs once uncertainty is resolved \citep{smith2019lagged,song2016experiment_on_ref}.\footnote{As argued by \cite{heffetz2021sinkin}, in experimental settings consistent with the KR model, a brief span as short as five minutes may be sufficient for reference point updates.} In line with this evidence, I assume that the consumer's reference is shaped by her most recent belief, formed at the beginning of each current stage.

In this paper, I explore the advance-purchase game when a consumer is loss-averse and bases her reference on her most recent beliefs. I apply the solution concept of ``Preferred Personal Equilibrium'' (PPE) proposed by \cite{KR09}, which posits that the consumer will follow her most favorable plan in each stage.

One of my key findings is that a consumer's loss aversion increases the price at which she is willing to pre-purchase. Interestingly, this price can even exceed the expected value of the good.
This finding indicates that the seller can profit from a consumer's loss aversion without having to introduce stochasticities to manipulate the consumer's reference, a strategy that has been discussed in previous studies \citep{heidhues2014sales,rosato2016selling}. This result can be attributed to two aspects of loss aversion: the fear of potentially paying a higher price in the future, and the aversion to the disappointment arising from not consuming the good.
Both types of losses arise from the consumer's concern about possible future outcomes, and are more severe if the consumer chooses to wait in the advance-purchase stage. 
Specifically, both aversion to consumption-value loss and aversion to monetary loss increase the consumer's willingness to pre-purchase.

Moreover, this study demonstrates that the seller has a strict incentive to commit to the spot price in advance even though he is not obliged to.\footnote{This commitment behavior is commonly observed and can be in various forms, such as the seller announcing ``the price was \$A, now you pay \$B"; or that ``the early-bird price is `x\% off' the regular price", allowing consumers to infer future prices.} This finding positions the seller's commitment as a strategic choice, endogenizing what the previous advance-purchase literature typically assumed. The underlying mechanism is related to the consumer's aversion to the loss in consumption value, as discussed earlier. The aversion to such a loss is mitigated if the seller does not commit to a future spot price. This is because the consumer will expect the seller to adjust the price downward to ensure that the good is sold. Consequently, a seller's decision not to commit to a spot price reduces the profit he could extract from this type of loss aversion.

This paper contributes to the literature on expectation-based reference-dependent preference by incorporating distinct sensitivities for different types of losses instead of a single parameter. This enables a more detailed understanding of which type of loss serves as the determinant of the results.

Furthermore, some existing applications use the static model proposed by \cite{kHoszegi2006model} to approach situations with sequential moves through ``static analysis''.\footnote{See also a comment by \cite{spiegler2012et} on using static models for dynamic problems.} This approach, while useful in simplifying certain contexts, assumes a fixed reference and overlooks the potential for reference updating \citep{herweg2013newsvendor,heidhues2014sales,herweg2015loss,karle2020selling,herweg2010binary,herweg2013flatrate,hahn2018et}.\footnote{However, the timing of reference determination varies in the literature. For instance, \cite{heidhues2014sales} assume the consumer forms her reference based on a belief at the root of a game, established before the seller offers a contract. Conversely, \cite{herweg2013flatrate} posits that the consumer's reference is formed after the seller's contract offer but before her first active stage.} Notably, the research of \cite{karle2020selling} also investigates advance selling. However, they primarily concentrate on the effects of improved consumer information at the advance-purchase phase, as evaluated through static analysis. In contrast, this paper introduces a novel perspective by incorporating reference updating. Section \ref{sec:fast_play} reveals a reverse effect of loss aversion when the reference is static. This reverse effect is relevant to the fact that the qualification of a Preferred Personal Equilibrium (PPE) is sensitive to the updating of the reference. This discussion suggests the potential insights that can be gained from considering reference updates in dynamic decision-making scenarios.

This paper also adds to the advance-purchase literature by endogenizing the seller’s price-commitment behavior, without relying on additional industry-specific assumptions that previous studies had made \citep{dana1998advance,shugan2000advance,xie2001advancesell,zhao2010pre,karle2020selling}. Therefore, it shows that the spot-price commitment, crucial for the profitability of advance sales as established in existing literature, can be deduced as a strategic choice of a profit-maximizing seller when faced with a loss-averse consumer. This approach retains the analysis in its simplest form without introducing additional assumptions, while simultaneously offering new insights into the seller's behavior.

A comparative analysis between scenarios with and without reference updating tentatively suggests that the time interval in the advance-purchase problem may significantly influence the behavior of a loss-averse consumer. Specifically, it is posited that an increase in the time interval between the resolution of uncertainty and the subsequent decision stages could ensure a more definitive update of the reference, thereby potentially solidifying the effect of loss aversion on boosting the consumer's willingness to pre-purchase. A monopolist who understands this influence could lengthen the interval to facilitate reference updating and thereby aiming to enhance profits. This line of reasoning provides an exploratory insight into the potential impact of reference updating, leaving further research to unravel its complexities and interactions.

Section \ref{sec:model} outlines the model, featuring the advance-purchase game form, consumer utility, and solution concept. Section \ref{sec:results} presents optimal pricing. Section \ref{sec:commitment} extends the advance-purchase game by endogenizing the seller's spot-price commitment behavior. Section \ref{sec:further} discusses two distinct scenarios: one in which the consumer is risk-averse, and the other in which the reference for a loss-averse consumer remains fixed. Section \ref{sec:conclusion} concludes remarks. Appendix A explores the role of the second trading opportunity within the advance-purchase problem. Appendix B contains the proofs. 

\section{Model}\label{sec:model}

\subsection{The Advance-purchase Game Form}

In this section, the \emph{advance-purchase game form} is introduced, involving two parties: a seller ($s$, he) and a representative consumer ($c$, she). 
The seller, a monopolist, has goods ready to sell. The consumer has unit demand. 
Assuming two states of nature, $\omega \in \{H,L\}$, where $H$ and $L$ represent high and low values of the good to the consumer, occurring with probabilities $q\in (0,1)$ and $1-q$ respectively, a chance player $0$ is introduced to randomly determine the state according to these probabilities.

The seller first offers a pair of prices $(p_1,p_2)\in \mathbb{R}_+^2$---the \emph{advance price} and the \emph{spot price}. Observing these prices, the consumer proceeds to the \emph{advance-purchase stage}, denoted by $T_1$, faced with the uncertainty about the state of nature. This uncertainty reflects the fact that the consumer's demand is influenced by currently unknown factors such as future weather conditions or psychological states. Let the initial expected valuation be $\mathbb{E}[\omega]:=q H + (1-q) L$, and let the consumption utility of not consuming be normalized to $0$. 
It is assumed that $H>L>0$, indicating that the consumer finds consuming the good more appealing than not consuming. In stage $T_1$, the consumer determines whether to \textit{pre-purchase} the good or to \textit{wait}. If the consumer chooses to pre-purchase, she pays the advance price, $p_1$, and is assured of receiving the good at the end of the interaction. Conversely, if she chooses to wait, an opportunity to buy the good in the \emph{spot-market stage} ($T_2$) arises. Before reaching this stage, the consumer realizes the value of the good. In the spot-market, she then decides whether to \emph{buy} the good at the spot price, $p_2$, or to \emph{reject} it. For convenience, the consumer is referred to as the $T_1$-consumer or the $T_2$-consumer based on the stage.

The timeline is summarized:
\begin{enumerate}
    \item The seller offers $(p_1,p_2)\in \mathbb{R}_+^2$.
    \item Upon observing the prices, the consumer progresses to $T_1$. She then decides either to \emph{pre-purchase} at $p_1$ or to \emph{wait}.
    \item The state of nature $\omega\in \{H,L\}$ is realized by the consumer.
    \item After realizing $\omega$, the consumer who did not pre-purchase enters $T_2$. She then decides whether to buy the good at the spot price, $p_2$.
    \item The good, once purchased, will be delivered at the end.
\end{enumerate}

The assumption that the seller commits to the spot price in advance is in line with previous advance-purchase literature.\footnote{See \cite{gale1992airticket,gale1993advance}, \cite{dana1998advance}, \cite{shugan2000advance}, \cite{prasad2011advance} and \cite{karle2020selling}. An exception is \cite{prasad2011advance}, which assumes the spot price is exogenously determined.}
This assumption is relaxed in Section \ref{sec:commitment}, where an alternative game is introduced. In that game form, the seller has the option to decide in advance whether to commit to $p_2$.

\begin{figure}[H]
    \centering
    \includegraphics[width=.8\textwidth]{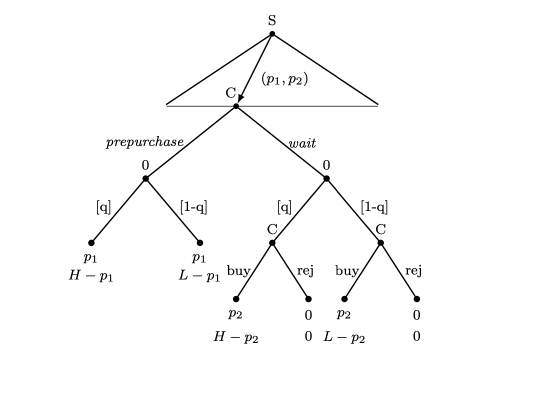}
    \caption{Game Tree}
    \label{fig:ori_game_tree}
\end{figure}

Figure \ref{fig:ori_game_tree} illustrates the game tree. Let $\mathcal{H}$ be the set of non-terminal nodes, partitioned into $\mathcal{H}_s$,\footnote{In the game tree shown in Fig. \ref{fig:ori_game_tree}, the set $\mathcal{H}_s$ contains only the root. This set is expanded to include more nodes in the extended version of Section 4.} $\mathcal{H}_c$, and $\mathcal{H}_0$, for the nodes where the seller, the consumer, and chance take action, respectively. The advance-purchase stage, $T_1$, consists of nodes of the form $h= (p_1,p_2)$ for any feasible prices $p_1, p_2\in \mathbb{R}_+$.  
The spot-market stage, $T_2$, consists of nodes of the form $h=\big( (p_1,p_2),wait,\omega \big)$ for all $p_1$, $p_2$, and $\omega\in \{H, L\}$, which indicates that the $T_2$-consumer takes action upon observing the state $\omega$. This setup implies that $\mathcal{H}_c=T_1\cup T_2$. For any two $h,h'\in  \mathcal{H}$, let $h\prec h'$ denote that $h$ precedes $h'$, illustrating the sequence of actions taken. For instance, $(p'_1,p'_2)\prec \big((p'_1,p'_2),wait,\omega\big)$.

The set of available actions for the consumer, $A_c(h)$, is given by:
\begin{itemize}
    \item at a node $h=(p_1,p_2)\in T_1$, $A_c(h)=\{prepurchase, wait\}$;
    \item at a node $h=\big( (p_1,p_2),wait,\omega \big) \in T_2$, $A_c(h)=\{buy, reject\}$.
\end{itemize}

Let $Z$ be the set containing all terminal nodes. The consumer's \emph{material payoff function}, $g_c:Z\to K\subseteq \mathbb{R}^2$, maps each terminal node to a vector-valued material payoff, denoted as $k=(k^V,k^M)\in K $. It consists of two dimensions: the (consumption) \emph{value dimension} ($k^V$) and the \emph{monetary dimension} ($k^M$). The former stems from the consumption utility, while the latter, which is non-positive, is derived from financial expenditures. On the other hand, the seller's \emph{material payoff} function, defined as $g_s:Z\to \mathbb{R}$, assigns a profit level to each terminal node.\footnote{Assume that the seller will discard the unsold good.} For instance, at the terminal node $z=\left((p_1,p_2), prepurchase,H\right)$, the consumer's material payoff is $g_c(z)=(H, -p_1)$ and the seller's material payoff is $g_s(z)=p_1$.

\subsection{Preference of a Loss-averse Consumer}
Having outlined the game form, I now define the utilities of each player in the advance-purchase game. In this game, the seller aims to maximize profit; therefore, $u_s(z)=g_s(z)$ for all $z\in Z$. The consumer's utility function, $u_c$, is based on the KR utility model, as outlined by the function form of \cite{KR09}. It incorporates both the material payoff and a \emph{loss utility}. Specifically, the consumer experiences a loss utility if her material payoff in either the value or the monetary dimension falls below the respective level of a two-dimensional reference point $r=(r^V,r^M) \in K$, consisting of a value dimension ($r^V$), and a monetary dimension ($r^M$). For a vector-valued material payoff $k$ and a reference point $r$, the loss-averse consumer's \emph{riskless utility} in dimension $d\in \{V,M\}$ is:

\begin{equation}
    u^d_c(k^d,r^d)=k^d- \lambda^d \cdot  \mathcal{L}(r^d-k^d). \label{eq:u_c^d}
\end{equation}

The loss utility function, $\mathcal{L}(x)=\max\{x,0\}$, is non-negative and exhibits a truncated linear form. Here, $\lambda^V\geq 0$ and $\lambda^M\geq 0$ are respectively the \emph{behavioral coefficients} of loss aversion in the value and monetary dimensions.\footnote{
The original KR utility function considers both gains and losses, with losses weighted more heavily to reflect "loss aversion." Conversely, my model focuses solely on the utility from losses, with $\lambda$'s representing the behavioral, or \emph{de facto}, coefficients of loss aversion. It can be shown It can be shown that on-path utility remains exchangeable regardless of whether gain utility is included, see, for example, \cite{dreyfuss2019expectations}. Similarly, studies such as \cite{herweg2013newsvendor}, \cite{herweg2015loss} apply similar assumptions to explore loss aversion’s impact on consumer decisions, focusing more on its presence than on the extent of utility asymmetry. \label{footnote_loss} }

When a deterministic reference point $r$ is fixed, the consumer's utility at a given terminal node is expressed as:
\begin{equation}
u_c(k,r):=u^V_c(k^V,r^V)+u^M_c(k^M,r^M)=k^V-\lambda^V\cdot \mathcal{L}(r^V-k^V) + k^M-\lambda^M\cdot \mathcal{L}(r^M-k^M). \label{eq:riskless_utility}
\end{equation}

This equation indicates the feature of narrow bracketing, a key assumption in the KR utility model, wherein losses are calculated independently in each dimension.
Given a reference point $r=(r^V,r^M)$, a consumer expects potential losses from two perspectives: (i) spending more than the reference (if $k^M<r^M$), and (ii) experiencing disappointment due to not achieving a certain level of consumption utility (if $k^V<r^V$). Different from previous applied literature, this paper permits the consumer to demonstrate distinct sensitivities toward different types of losses, namely $\lambda^V\neq \lambda^M$. This differentiation enables the current approach to examine the impact of loss aversion in each dimension on the results.\footnote{Experimental evidence suggests that the coefficient of loss aversion in good and money can be different \citep[p. 326]{dellavigna2009psychology}.}

While \eqref{eq:riskless_utility} applies when a deterministic reference point is specified, \cite{kHoszegi2006model,KR09} propose that in environments with uncertainty, the player's reference is not fixed but is determined by her distributional beliefs---a concept referred to as a stochastic reference point. To distinguish it from the previous notation of a reference point, $r$, it will be hereafter referred to as a \emph{reference distribution}. 
As the advance-purchase game unfolds, the consumer can potentially make two moves, with her beliefs being updated in response to new information. Therefore, in this dynamic game, it becomes critical to specify how the reference distribution is determined in each stage.

A crucial assumption of this approach is that the consumer's decision-making is influenced by her \emph{recent belief}.\footnote{The concept of the recent-belief reference distribution is akin to the notion of ``slow play'' discussed in some studies. For example, in the working paper version of \cite{BDS_working_paper} investigating anger, slow play refers to adjusting the reference expectation closer to the moment of action, as opposed to ``fast play,'' where the reference expectation remains fixed at the start of the game. From another perspective, the recent-belief reference distribution in slow play can also be seen as a form of ``fast adjustment'', indicating that it updates rapidly \citep{smith2019lagged,Thakral2021}.} This concept is formalized:

\begin{assumption}
 The consumer's reference distribution in each decision stage is determined by her recent belief, which is the latest updated belief at the beginning of that stage.
    \label{assu:recent_belief}
\end{assumption}

To formally define a reference distribution, I now establish the consumer's (first-order) belief system as a conditional probability system, assigning probabilities to terminal nodes, or equivalently, paths $z\in Z$ at each non-terminal node $h \in \mathcal{H}$. Let $Z(h)\subseteq Z$ denote the set of all possible paths through $h$. Each probability, $\alpha_c(z|h)\in \Delta (Z(h))$, represents the conditional belief at $h$ that a path $z\in Z(h)$ will be reached. For convenience, when $h$ belongs to $\mathcal{H}_c$, indicating the consumer's turn to move, I use $\alpha_{c,c}(\cdot|h)$ to represent her belief about the action $a_c\in A_c(h)$ she can take. For instance, at $h=(p_1,p_2)\in T_1$, $\alpha_{c,c}(wait|h)$ represents her probabilistic belief in taking the ``wait'' action upon observing the prices $(p_1,p_2)$.\footnote{Formally, $\alpha_{c,c}(wait|h)$ is defined as $\alpha_{c,c}(wait|h):= \alpha_c(Z(h')|h)$, where $h'=(h,wait)$ represents the node immediately following $h$ when the consumer takes the ``wait'' action.} 
The consumer's system of \textbf{first-order beliefs} is defined as $\alpha_c:=\left(\alpha_c(\cdot|h)\right)_{h\in \mathcal{H}}$. This system is a conditional probabilistic system that satisfies the \emph{consistency} condition. Specifically, for any pair of $h,h'\in \mathcal{H}: h\prec h'$ and $Y\subseteq Z(h')$,

\begin{equation}
    \alpha_c\left( Z(h') |h \right)>0 \Rightarrow \alpha_c\left(  Y|h'\right) =\frac{\alpha_c\left( Y|h\right)  }{\alpha_c\left(  Z(h')|h\right) } \label{condition:consistent_belief}
\end{equation}

In this game, a path $z$ captures the series of actions taken by all players, including the seller, chance, and the consumer herself. Consequently, at each node $h\in \mathcal{H}$, the consumer's belief, $\alpha_c(z|h)$, reflects her probabilistic belief about which path $z$ will materialize, and thus about the sequence of future actions that each path $z$ encompasses. 
Accordingly, the belief system $\alpha_c$, which contains all such conditional beliefs as $\left(\alpha_c(\cdot|h)\right)_{h\in \mathcal{H}_c}$, can be divided into two distinct components---her beliefs about her own actions at all of her decision nodes $h\in \mathcal{H}_c$, forming her \textbf{global plan} denoted by $\alpha_{c,c}\in \times_{h\in \mathcal{H}_c} \Delta\big(A_c(h)\big)$, and her beliefs about the behavior of others, including the chance player, represented as $\alpha_{c,-c}\in \times_{h\in \mathcal{H}_{s}\cup \mathcal{H}_0}\Delta\big(A_{j}(h)\big)$, where $j$ is the active player at $h$.\footnote{Cf. \cite{BCD,jelpgt}.}\footnote{Analogous to $\alpha_{c,c}(\cdot|\cdot)$, $\alpha_{c,-c}(a_{-c|h})$ is defined as $\alpha_{c,-c}(a_{-c|h}):=\alpha_c(Z(h')|h)$ where $h'=(h,a_{-c})$ is the node immediately following $h$ when the action $a_{-c}$ is taken by others.}
In the advance-purchase game, the consumer's conditional beliefs at a node in the first stage, $T_1$, about her onward actions in both stages, are defined as her \textbf{plan}. The global plan, $\alpha_{c,c}$, is thus, an aggregate of all such plans contingents on any possible pairs of prices $(p_1,p_2)\in \mathbb{R}^2_{+}$, and it functions equivalently as a behavior strategy.
As the game progresses to the second stage, $T_2$, her conditional beliefs at the corresponding node in this stage about her immediate purchasing behavior in the spot market, are referred to as her \textbf{sub-plans}. 
Condition \eqref{condition:consistent_belief} indicates that the sub-plan of the $T_2$-consumer should be correctly integrated into the plan of the $T_1$-consumer, ensuring that the latter thoughtfully encompasses the former for the second stage.\footnote{Formally, this condition ensures that for any pair of nodes $h\in T_1$ and $h'\in T_2$ where $h\prec h'$, and for any action $a_c\in A_c(h'), \alpha_c\left(Z(a_c)|h'\right)=\left(Z(h')|h\right)\cdot \left(Z(h',a_c)|h\right)$. This alignment ensures that the sub-plan at $h'$ is effectively a part of her (broader) plan at $h$, reflecting that her beliefs are consistent across stages.} 
Since the consumer's decision stages occur after she observes the prices, the following discussion will focus primarily on her ``plans'' or ``sub-plans'' in the advance-purchase game, detailing her beliefs about subsequent actions, formulated after she observes prices. However, when discussing solution concepts that require a comprehensive description of the consumer's responses contingent on different prices, the broader concept of ``global plan'' will be used.

The consumer's \textbf{reference distribution} is a probabilistic belief about material payoffs, i.e., an element in $\Delta (K)$. According to Assumption \ref{assu:recent_belief}, at each decision node $h\in \mathcal{H}_c$, this belief is specified as to her up-to-date belief, $\alpha_c(\cdot |h)$.  Furthermore, in line with the model of \cite{kHoszegi2006model,KR09}, the reference distribution is fixed while the consumer evaluates different actions.\footnote{In another paper, \cite{KR2007risk} considered an alternative concept of a ``choice-acclimating personal equilibrium'' (CPE), in which the player chooses both her reference distribution and her action simultaneously.}

The consumer's expected utility for taking an action $a_c\in A_c(h)$ is:
\begin{equation}
     \begin{array}{lll}
        \Bar{u}_{c,h}(a_c,\alpha_c(\cdot|h))
        &:= \mathbb{E}_{\alpha_c} \left\{ \sum \limits_{\tilde{z}\in Z(h)} \alpha_c(\tilde{z}|h) \cdot u_c(g_c(\cdot) ,g_c(\tilde{z})) \bigg|h,a_c
        \right \}\\
         &= \sum \limits_{z\in Z(h,a_c)}{\alpha_c(z|h,a_c)} \cdot \left\{ \sum \limits_{\tilde{z}\in Z(h)}{\alpha_c(\tilde{z}|h)}\cdot \big[ u_c\big(g_c(z), g_c(\tilde{z})\big) \big]  \right\} .
    \end{array}
     \label{eq:U_function}
\end{equation}

The equation encapsulated within the braces is the consumer's utility given a material payoff $k\in K$, and a reference distribution induced by her \emph{recent belief}, $\alpha_c(\tilde{z}|h)$.\footnote{
Equation \eqref{eq:U_function} can be restated using the notation by \cite{KR09}: Let $F_{h}$ and $F_{h,a_c}$ denote the distributions over the set of the consumer's material payoff, $K$, resulting from the consumer's belief conditional on node $h$ and $(h,a_c)$ respectively. Following the framework of \citeauthor{KR09}, the expected utility function, derived from the action $a_c$, is expressed as $U_h(F_{h,a_c},F_h)$. The first argument, $F_{h,a_c}$, is the material payoff lottery ensuing from the action $a_c$. The second term, $F_h$, is the reference distribution, which is fixed when the consumer considers taking action $a_c$.
} 
In a specific dimension, $d\in \{V,M\}$, this equation indicates that a loss-averse consumer experiences mixed loss sensations when comparing a material payoff level, $k^d$, against different reference points, $r^d$, on the support of the reference distribution \citep{kHoszegi2006model,KR09}. The full specification of these expected utilities in the advance-purchase game is elaborated in Section \ref{sec:main_loss_averse_consumer_pre-purchasing}.

Readers familiar with the model of \cite{KR09} will notice that Assumption \ref{assu:recent_belief} used in this paper differs from theirs. \cite{KR09} posits that a player's reference distribution is determined by her \emph{lagged beliefs}—the belief updated in the previous period. Experimental studies by \cite{smith2019lagged} and \cite{song2016experiment_on_ref} suggest that players' references update rapidly, accommodating new information. This finding does not support the lagged-belief assumption.\footnote{\cite{heffetz2021sinkin} suggests that this procedure also depends on whether the new information has sunk in.} In light of this evidence, the analysis and discussions hereafter in this paper will be grounded in Assumption \ref{assu:recent_belief}.

The consumer's KR utility depends on her own beliefs, particularly, her belief-based reference distribution. This places the advance-purchase game within the framework of the ``psychological game theory''\citep{gps,dgp,jelpgt}. Unlike traditional game theory, which primarily focuses on players' beliefs about co-players' actions, the advance-purchase game also focuses on what a player believes about actions that she is going to take. Here, a plan affects the expected utility in two ways---it indirectly affects the expected utility by outlining the probability of outcomes, and it also directly affects the utility function through the reference distribution.
As a result of such a direct impact of the plan, the loss-averse consumer's preference exhibits dynamic inconsistency.\footnote{See also the discussions of own-plan dependence by \cite{BCD,jelpgt}, and \cite{kHoszegi2006model,KR09}.} To address this inconsistency, the forthcoming subsection introduces the solution concept of ``personal equilibrium'', as termed by \cite{kHoszegi2006model,KR09}, which satisfies the fixed-point condition.

\subsection{Solution Concepts for Loss-averse Consumer Behavior}

To maintain consistency between the actions and beliefs of a loss-averse consumer across stages, \cite{kHoszegi2006model} imposed the solution concept of the \emph{personal equilibrium}.

\begin{defn}
    A global plan $\alpha_{c,c}^{\ast}$ is a \textbf{Personal Equilibrium} (PE) if, for any node $h\in T_1\cup T_2$,
    \begin{equation}
        \alpha_{c,c}^{\ast}(a_c^{\star}|h)>0 \Rightarrow a_c^{\star} \in \argmax_{a_c\in A_c(h)} \Bar{u}_{c,h}(a_c,\alpha_c^{\ast}(\cdot|h)) \label{eq_def:PE} 
    \end{equation}
    where $\alpha_{c,c}^{\ast}$, the global plan, and $\alpha_c^{\ast}(\cdot|h)$, the conditional beliefs at node $h$, are both derived from the same first-order belief system $\alpha_c^{\ast}$.
\end{defn}

A global plan is deemed \emph{credible at $h$} if it satisfies the condition in \eqref{eq_def:PE} at that node.
This condition ensures that the action prescribed by the global plan is \emph{locally optimal}, meaning that the loss-averse consumer has no strict incentive to deviate from this action at $h$. The set of all such global plans that are credible at $h$ is denoted by $P(h)$.

A personal equilibrium is established as credible at the corresponding decision nodes within $\mathcal{H}_c= T_1\cup T_2$, meaning that $\alpha^{\ast}_{c,c} \in \bigcap_{h\in T_1\cup T_2} P(h)$. This implies that only actions that are locally optimal, given that the reference distribution is based on the belief consistent with the plan, are assigned positive probabilities by the plan.\footnote{\cite{jelpgt} refer to this property as ``rational planning''.} In this context, for any pair of nodes $h\in T_1$ and $h'\in T_2$ where $h\preceq h'$, the condition $\alpha_{c,c}^{\ast}\in P(h)\cap P(h')$ demonstrates the sophistication of the consumer. Specifically, in stage $T_1$, the consumer considers only plans that are consistent with the sub-plans deemed credible at each node in stage $T_2$. 

Given that a loss-averse consumer may encounter multiple personal equilibria, the concept of ``Preferred Personal Equilibrium'' (PPE) was introduced by \cite{KR09} to refine the solution concept in dynamic decision environments. A PPE, in the advance-purchase game, aligns with the most favorable sub-plans credible at the corresponding nodes in $T_2$ and maximizes the consumer's expected utility in $T_1$, for any pair of prices.

Reflecting the sophistication inherent in a personal equilibrium, a consumer adhering to a PPE considers, when in $T_1$, only the most favorable sub-plan that is credible at the node in $T_2$, even if it does not maximize her expected utility in $T_1$. Thus, the PPE is constructed through a backward process, ensuring this consistency across stages.

\begin{defn}
A global plan $\alpha_{c,c}^{\ast\ast}$ is a \textbf{Preferred Personal Equilibrium} (PPE) if it satisfies the following conditions:
\begin{itemize}
    \item[(i)] For each node $h^{\prime}\in T_2$, each global plan $\tilde{\alpha}_{c,c}$ belonging to $ P(h^{\prime})$, and each action $a_c^{\prime}$ with $\alpha_{c,c}^{\ast\ast}(a_c^{\prime}|h^{\prime})>0$ and $\tilde{a}_c$ with $\tilde{\alpha}_{c,c}(\tilde{a}_c|h^{\prime})>0$, the following condition holds: 
  
    \[\Bar{u}_{c,h^{\prime}}(a_c^{\prime},\alpha_c^{\ast\ast}(\cdot|h^{\prime}))\geq \Bar{u}_{c, h^{\prime}}(\tilde{a}_c,\tilde{\alpha}_c(\cdot|h^{\prime}));\]

    \item[(ii)] for any $h\in T_1$, $\alpha_{c,c}^{\ast\ast}$ satisfies that \[\alpha_{c,c}^{\ast\ast}(a_c^{\star}|h)>0 \Rightarrow a_c^{\star} \in \argmax_{a_c\in A_c(h)} \Bar{u}_{c,h}(a_c,\alpha_c^{\ast\ast}(\cdot|h)); \]

    \item[(iii)] for any $h\in T_1$, $\alpha_{c,c}^{\ast\ast}$ maximizes the consumer's expected utility conditional on $h$ among all the global plans that satisfy conditions (i) and (ii).

\end{itemize}
Note: both $\alpha_{c,c}^{\ast\ast}$ and $\alpha_c^{\ast\ast}(\cdot|h)$, as well as $\tilde{\alpha}_{c,c}$  and $\tilde{\alpha}_{c}(\cdot|h)$ are derived from their first-order belief systems $\alpha_c^{\ast\ast}$ and $\tilde{\alpha}_{c}$, respectively.
\label{defn:PPE}
\end{defn}

Condition (i) states that at any node $h^{\prime}\in T_2$, the sub-plan $\alpha_{c}^{\ast\ast}(\cdot|h^{\prime})$, derived from $\alpha_{c,c}^{\ast\ast}$, is the $T_2$-consumer's most favorable sub-plan among those that are credible at that node. In essence, this sub-plan is a preferred personal equilibrium within the sub-game starting at node $h^{\prime}$. Condition (ii) ensures that the $T_1$-consumer, whose reference distribution is shaped by correct beliefs about subsequent actions, has no incentive to deviate from the prescribed action by the plan $\alpha_{c,c}^{\ast\ast}(\cdot|h)$.  Finally, condition (iii) asserts that this plan is the most favorable for the $T_1$-consumer among those that satisfy conditions (i) and (ii).

Under Assumption \ref{assu:recent_belief}, which posits that the consumer's reference distribution reflects her recent beliefs, condition (i) suggests that the $T_2$-consumer can flexibly revise her sub-plan at the beginning of $T_2$ without incurring costs.
This revision allows her to implement the most favorable sub-plan, based on current preferences, among those that are credible at that node, regardless of the $T_1$-consumer's preference. Anticipating this flexibility, the $T_1$-consumer considers only plans that the $T_2$-consumer would be willing to implement without revision.\footnote{Contrary to my model, \cite{KR09}'s lagged-belief assumption allows the $T_1$-consumer to set a reference distribution for the $T_2$-consumer based on her own preference, making deviations psychologically costly for the $T_2$-consumer. Thanks to an anonymous reviewer for pointing out these implied psychological costs.}

\section{Consumer Behavior and Optimal Pricing}\label{sec:results}

This section examines consumer behavior and the seller's optimal pricing policy, assuming that the consumer adheres to the preferred personal equilibrium. In addition, for simplicity, the consumer is assumed to make a purchase whenever she is indifferent.\footnote{This assumption is not crucial for the equilibrium analysis of the strategic interaction, since if the consumer does not buy, the seller would deviate by lowering the price, leading to the current strategy profile to be non-equilibrium.}

\subsection{Benchmark: A Consumer with Standard Risk-neutral Preference}

If $\lambda^M=\lambda^V=0$, the consumer's utility function represents a standard risk-neutral preference. 
In the forthcoming sections involving standard preference, I apply the solution concept of sub-game perfect equilibrium.

\begin{remark}\label{rmk:2ndstage}
Given the pricing pair $(p_1,p_2)$, if $\lambda^M=\lambda^V=0$, then a consumer who has not pre-purchased will choose to buy whenever $p_2\leq \omega$ in the spot-market stage.
\end{remark}

When encountering such a consumer, the seller's optimal pricing extracts the full consumer surplus.

\begin{proposition}\label{prop:standard}
    If the consumer $\lambda^M=\lambda^V=0$, the seller's maximum profit equals $\mathbb{E}[\omega]$.
\end{proposition}

For example, when the seller sets $(p_1, p_2)=(\mathbb{E}[\omega],H)$, the consumer is willing to pre-purchase, allowing the seller to fully obtain the consumer surplus and achieve maximum profit. 
Furthermore, even a modest aversion to risk lowers the consumer's willingness to pre-purchase, thus reducing the cutoff advance price. This will be discussed in Section \ref{sec:RA}.

\subsection{A Consumer with Belief-based Loss-averse Preference}\label{sec:main_loss_averse_consumer_pre-purchasing}

In this subsection, I analyze the effect of consumer's loss aversion in an advance-purchase game, assuming that the consumer adopts a Preferred Personal Equilibrium (PPE).

\begin{lemma}\label{lemma:2ndstage}
A consumer who did not pre-purchase will choose to buy at a spot price that is not greater than the realized value, i.e., $p_2\leq \omega$.
\end{lemma}

This lemma implies that the consumer's highest acceptable spot price is independent of $\lambda^V$ or $\lambda^M$. Under Assumption \ref{assu:recent_belief}, neither the previous reference distribution nor the previous belief has any impact on the behavior of the $T_2$-consumer. Consequently, $T_1$-consumer cannot influence the $T_2$-consumer's behavior through her reference distribution. After learning $\omega$, the $T_2$-consumer behaves under a deterministic environment. In this situation, a $T_2$-consumer's who implements a PPE behaves identically to that of a consumer with a standard risk-neutral preference.\footnote{See also the shoe-shopping example in Section IV of \cite{kHoszegi2006model}.} 

To demonstrate that the $T_1$-consumer's influence on the $T_2$-consumer does not extend through the reference distribution, and that the $T_2$-consumer behaves under a deterministic environment in this context, consider a situation where $p_2$ falls in the range $(L, H)$, and the consumer has a (degenerate) plan. In this plan, the $T_1$-consumer pre-purchases, and the $T_2$-consumer, if active, chooses to buy at $h=((p_1,p_2), wait, H)$ and to reject at $h'=((p_1,p_2), wait, L)$. Suppose that the consumer arrives at $h=((p_1,p_2), wait,H)$. This implies that the $T_1$-consumer's action is inconsistent with her plan. The $T_2$-consumer must then update her reference distribution based on her belief condition on $h$. Even if the $T_1$-consumer's action deviates from her plan, the consumer is not considered to be revising her plan as long as, at $h$, the $T_2$-consumer's sub-plan is ``to buy'', aligning with the original plan. For the $T_2$-consumer, the expected utility of choosing to buy is $\Bar{u}_{c,h}(buy,\alpha_c(\cdot|h))=H-p_2>0$, while the utility of deviating to reject is $\Bar{u}_{c,h}(reject,\alpha_c(\cdot|h))=0-\lambda^V H <0$.\footnote{Under an alternative assumption such as the lagged-belief reference distribution by \cite{KR09}, the plan would result in a lottery $(H, q; L)$ in value and $-p_1$ in monetary dimensions as the reference distribution at $h$. Accordingly, the expected utility calculations at $h$ would be different: for choosing to buy, it becomes $H-p_2-\lambda^M \mathcal{L}(p_2-p_1)$, and for choosing to reject, it is $0-\lambda^V q H - \lambda^V (1-q) L$. Under these conditions, the statement of Lemma \ref{lemma:2ndstage} does not hold.}

Now the analysis turns to the consumer's behavior at $T_1$. The range of plans considered by the $T_1$-consumer is constrained to those aligning with the behavior patterns established in Lemma \ref{lemma:2ndstage}.

Moving forward with this analysis, three different sub-cases are considered based on the value of $p_2$. Consistent with the previous assumption that the consumer will opt to purchase when indifferent, I examine two classes of degenerate plans for each sub-case: one prescribes pre-purchasing at $T_1$ and the other prescribes waiting.
Note that each plan prescribes not only the immediate action at $T_1$, but also the contingent action at $T_2$ regardless of whether it is ultimately reached.\footnote{The rationale for considering only degenerate plans is that non-degenerate personal equilibria are not preferable to a degenerate one: They increase the fluctuation of the reference distribution, and hence the anticipatory losses.}

\subsubsection*{Sub-case $p_2\leq L$:}

In the scenario, the $T_2$-consumer will choose to buy the good at $p_2\leq L$ regardless of the state $\omega\in \{H,L\}$. Anticipating this behavior, the $T_1$-consumer's plan, whether to pre-purchase or to wait at $T_1$, must align with this anticipated action of the $T_2$-consumer. Accordingly, two distinct degenerate plans are considered. 

\paragraph{Plan to pre-purchase}

When formulating a plan to pre-purchase, the $T_1$-consumer expects a consumption utility of $H$ with probability $q$, and of $L$ with probability $1-q$, along with a certain monetary payoff of $-p_1$. These expectations form her reference distribution. 
Depending on this reference distribution, the $T_1$-consumer's expected utility for choosing to pre-purchase, derived from the general utility function in \eqref{eq:U_function}, is:
\begin{equation}
\begin{array}{ll}
    & q\cdot \Big\{q[H-\lambda^V\mathcal{L}(H-H)] + (1-q)[H-\lambda^V\mathcal{L}(L-H)] - p_1 -\lambda^M\mathcal{L}(-p_1-(-p_1)) \Big \} \vspace{1ex}\\
      + &(1-q)\cdot \Big\{q[L-\lambda^V\mathcal{L}(H-L)] + (1-q)[L-\lambda^V\mathcal{L}(L-L)] - p_1 -\lambda^M\mathcal{L}(-p_1-(-p_1)) \Big\} \vspace{1ex}\\
     = & qH + (1-q)L -p_1 -(1-q)\cdot q \cdot \lambda^V(H-L) . 
\end{array}\label{eq:implement_plan_pp}
\end{equation}
However, if the $T_1$-consumer deviates to wait (and then she will buy the good at $T_2$), her expected utility changes to be:
\begin{equation}
\begin{array}{ll}
     &  q\cdot \Big\{q[H-\lambda^V\mathcal{L}(H-H)] + (1-q)[H-\lambda^V\mathcal{L}(L-H)] - p_2 -\lambda^M\mathcal{L}(-p_1-(-p_2)) \Big\} \vspace{1ex}\\
       +& (1-q)\cdot \Big\{q[L-\lambda^V\mathcal{L}(H-L)] + (1-q)[L-\lambda^V\mathcal{L}(L-L)] - p_2 -\lambda^M\mathcal{L}(-p_1-(-p_2)) \Big\} \vspace{1ex}\\
     = & qH + (1-q)L -p_2 -(1-q)\cdot q \cdot \lambda^V (H-L) - \lambda^M \mathcal{L} (p_2-p_1),
\end{array}\label{eq:p2_low_pp_wait}
\end{equation}
where
\begin{equation*}
\mathcal{L}(p_2-p_1)=
\begin{cases}
p_2-p_1 & \text{if } p_2 \geq p_1,\\
0 & \text{otherwise.}
\end{cases}
\end{equation*}

Given these expected utilities, implementing the pre-purchasing plan is deemed locally optimal if the expected utility specified in \eqref{eq:implement_plan_pp} exceeds that of deviating to wait, detailed in \eqref{eq:p2_low_pp_wait}.

\paragraph{Plan to wait}

As this plan must align with the $T_2$-consumer's behavior to buy in either state $\omega\in \{H,L\}$, the $T_1$-consumer anticipates a consumption utility of $H$ with probability $q$, and $L$ with probability $1-q$, along with a certain monetary payoff of $-p_2$. These expectations form her reference distribution.

The $T_1$-consumer's expected utility for implementing the plan to wait is:
\begin{equation}
\begin{array}{ll}
     &  q\cdot  \Big \{q [ H -\lambda^V \mathcal{L}(H-H) ] +(1-q) [H-\lambda^V \mathcal{L} (L-H)] -  p_2 -\lambda^M \mathcal{L}(-p_2-(-p_2)) \Big \} \\
     + & (1-q)\cdot \Big \{q[L-\lambda^V\cdot \mathcal{L}(H-L)] +(1-q)[L-\lambda^V \cdot \mathcal{L}(L-L)  -p_2 -\lambda^M \mathcal{L}(-p_2-(-p_2))] \Big \} \\
      =& q\cdot H+(1-q)L-p_2-(1-q)q\lambda^V(H-L). \label{eq:implement_plan_wait}
\end{array}
\end{equation}

Conversely, if the she deviates to pre-purchase, her expected utility is:
\begin{equation}
\begin{array}{ll}
    &q \Big \{ q[H -\lambda^V \mathcal{L} (H-H)] +(1-q) [H-\lambda^V \mathcal{L} (L-H)]    -p_1 -\lambda^M\mathcal{L}(-p_2-(-p_1))] \Big \} \\
      + & (1-q) \Big\{q\cdot [L-\lambda^V(H-L)] +(1-q) [L -\lambda^V \mathcal{L} (L-L)]  -p_1-\lambda^M\mathcal{L}(-p_2-(-p_1))] \Big\}\\
      =& q\cdot H+(1-q)L-p_1-(1-q)q\lambda^V(H-L) -\lambda^M\mathcal{L}(p_1-p_2). \label{eq:deviate_plan_wait_to_pp}
\end{array}
\end{equation}

As with the pre-purchase plan, the waiting plan is deemed locally optimal if the expected utility, as in \eqref{eq:implement_plan_wait}, exceeds that of deviating to pre-purchase, detailed in \eqref{eq:deviate_plan_wait_to_pp}.

\subsubsection*{Sub-case $L<p_2\leq H$:}
In the scenario, the $T_2$-consumer will choose to buy the good if the realized state is $\omega=H$.

\paragraph{Plan to pre-purchase}

The reference distribution for this plan is identical to that of the previous sub-case, leading to the same expected utility for choosing to pre-purchase, as detailed in \eqref{eq:implement_plan_pp}. In contrast, if the $T_1$-consumer deviates to wait (and then buy at $T_2$ whenever $\omega=H$), her expected utility changes to be:
\begin{equation}
\begin{array}{ll}
     &  q\cdot \Big\{q[H-\lambda^V\mathcal{L}(H-H)] + (1-q)[H-\lambda^V\mathcal{L}(L-H)] - p_2 -\lambda^M\mathcal{L}(-p_1-(-p_2)) \Big\} \vspace{1ex}\\
      + & (1-q)\cdot \Big\{q[0-\lambda^V\mathcal{L}(H-0)] + (1-q)[0-\lambda^V\mathcal{L}(L-0)] -\lambda^M\mathcal{L}(-p_1-0)\Big \} \vspace{1ex}\\
      = & q\cdot \big\{H - p_2 - \lambda^M\mathcal{L}(p_2-p_1) \big\} + (1-q)\cdot \big\{-q\lambda^V\cdot H - (1-q)\lambda^V\cdot L \big \}.
\end{array}
\end{equation}

\paragraph{Plan to wait} 

In this sub-case, the $T_1$-consumer's plan to wait must be consistent with the $T_2$-consumer's behavior to buy only if $\omega=H$. Anticipating this behavior, she expects a consumption utility of $H$ with probability $q$, and $0$ with probability $1-q$. Accordingly, her expected monetary payoff is $-p_2$ with probability $q$, and $0$ with probability $1-q$. These expectations form her reference distribution.

The $T_1$-consumer's expected utility for choosing to wait is
\begin{equation}
    \begin{array}{ll}
        & q\cdot \Big \{ q[ H - \lambda^V\mathcal{L}(H-H)]+(1-q)[H-\lambda^V \mathcal{L} (0-H)]\\
        +& q[ -p_2 - \lambda^M \mathcal{L} (-p_2-(-p_2)) ]
        +(1-q)[-p_2-\lambda^M (0-(-p_2)) \Big\} \\
        + &(1-q)\cdot  \Big \{q \cdot [0-\lambda^V \mathcal{L}(H-0)]+(1-q) [0 -\lambda^V \mathcal{L} (0-0)  ]   \\
        +& q[ 0 - \lambda^M \mathcal{L} (-p_2-0) ]
        +(1-q)[0-\lambda^M \mathcal{L}(0-0) \Big\} \\
         = & q\cdot \big\{ H-p_2-(1-q)\lambda^M \cdot p_2\} + (1-q) \cdot \{q \cdot [0-\lambda^V\cdot H] \big\}.
    \end{array}
\end{equation}

However, if she deviates to pre-purchase, her expected utility then becomes:
\begin{equation}
\begin{array}{ll}
& q\cdot \Big\{ q[H-\lambda^V \mathcal{L}(H-H)]+(1-q) [H -\lambda^V \mathcal{L}(0-H)] \\
 +& q[ - p_1  -  \lambda^M\mathcal{L}(-p_2-(-p_1))] + (1-q)[-p_1 - \lambda^M \mathcal{L} (0-(-p_1)) \Big\}\\
     +&  (1-q)\cdot \Big \{ q\cdot [L-\lambda^V\mathcal{L}(H-L)] + (1-q)\cdot [L -\lambda^V\mathcal{L}(0-L)] \\
     +& q[ - p_1  -  \lambda^M\mathcal{L}(-p_2-(-p_1))] + (1-q)[-p_1 - \lambda^M \mathcal{L} (0-(-p_1)) \Big\}\\
     =& q\cdot H +(1-q)L-p_1 - (1-q)q\lambda^V (H-L)-q\lambda^M\mathcal{L}(p_1-p_2)-(1-q)\lambda^M\cdot p_1.
\end{array}
\end{equation}

\subsubsection*{Sub-case $p_2>H$:}
In this sub-case, the $T_2$-consumer will not buy regardless of the realized state $\omega\in \{H,L\}$.  Consequently, for the $T_1$-consumer, waiting at $T_1$ is effectively equivalent to choosing never to purchase the good.

\paragraph{Plan to pre-purchase}

Since the reference distribution remains identical to earlier sub-cases, the expected utility for implementing the pre-purchasing plan is again given by \eqref{eq:implement_plan_pp}. However, if the $T_1$-consumer deviates to wait, her expected utility changes to be:
\begin{equation}
\begin{array}{ll}
     &  q\cdot  [0- \lambda^V \mathcal{L}(H-0)] +(1-q) \cdot [0- \lambda^V \mathcal{L}(L-0) ] -0 -\lambda^M \mathcal{L}(-p_1-0)\\
     =&  0 -q \cdot \lambda^V \cdot H -(1-q) \cdot \lambda^V \cdot L .
\end{array}
\end{equation}

\vspace{-1em}

\paragraph{Plan to wait} 

The $T_1$-consumer's plan to wait must incorporate her never-purchase behavior at $T_2$. Accordingly, she expects a consumption utility of $0$, along with a monetary payoff of $0$ with certainty. 

The $T_1$-consumer's expected utility for implementing the plan to wait is 
\begin{equation}   
      0-\lambda^V \cdot \mathcal{L}(0-0) -\lambda^M \cdot \mathcal{L} (0-0) =0 . \label{eq:p2_high_wait_wait}
\end{equation}
If she instead deviates to pre-purchase, the expected utility is then:
\begin{equation}   
      \begin{array}{ll}
           & q \cdot \Big\{H-\lambda^V \cdot \mathcal{L} (0-H) -p_1 -\lambda^M\cdot \mathcal{L} (0-(-p_1)) \Big\} \\
           &+ (1-q) \cdot \Big \{L-\lambda^V \cdot \mathcal{L} (0-L) -p_1 -\lambda^M\cdot \mathcal{L} (0-(-p_1)) \Big\}  \\
           =& q \cdot H + (1-q) \cdot L -\lambda^M\cdot  p_1 .
      \end{array}\label{eq:p2_high_wait_deviate_pp}
\end{equation}

After establishing the expected utilities for each sub-case, I proceed to evaluate the conditions under which a plan is consistent with a personal equilibrium (PE). This happens if any deviation from it results in a weakly lower expected utility, which echoes the ``locally optimal'' concept as initially discussed in the first sub-case. For example, in the $p_2 > H $ sub-case, consider a plan that assigns the consumer to wait at $T_1$. This plan is consistent with a PE if the expected utility from implementing, which equals $0$, is weakly greater than the expected utility from deviating from it, as detailed in \eqref{eq:p2_high_wait_deviate_pp}.

Furthermore, in situations where multiple personal equilibria exist, the preferred personal equilibrium (PPE) is the one that yields the highest expected utility. Consider again the sub-case $p_2 > H $. Suppose both a plan that assigns the consumer to pre-purchase and a plan that assigns the consumer to wait are PEs. The plan to wait is consistent with a PPE requires that its expected utility, which equals $0$, is weakly greater than that of the plan to pre-purchase, $qH+(1-q)L-p_1-(1-q)\cdot q\cdot \lambda^V(H-L)$.

\begin{figure}[H]
	\centering
	\includegraphics[height=.5\textheight]{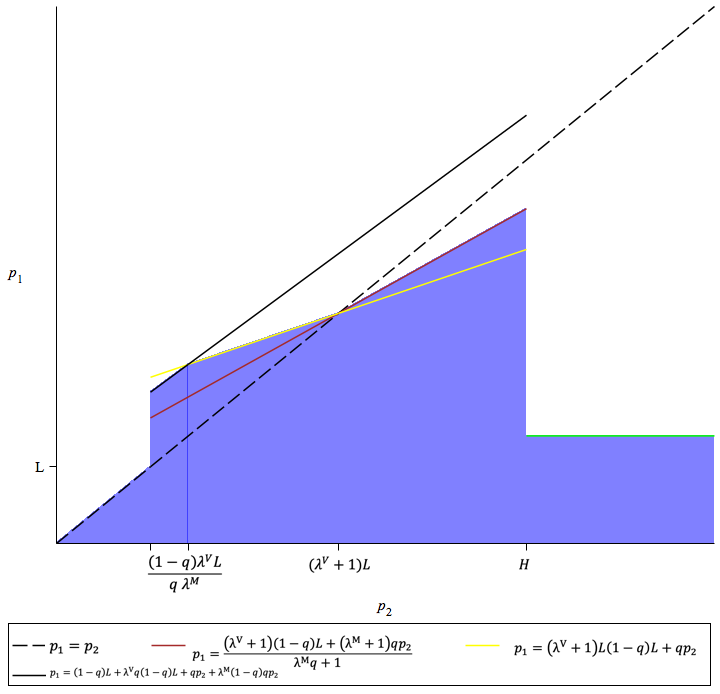}
	\caption{pre-purchasing at $p_1$---a case when $\frac{1-q}{q}\frac{\lambda^V}{\lambda^M}L<(\lambda^V+1)L<H$}
	\label{fig:PPPPE}
\end{figure}

Given the consumer follows a PPE, the  \emph{cutoff} advance price, $\hat{p}_1(p_2)$, below which she will buy, is characterized:

\begin{lemma}\label{lemma:monotone}
The cutoff price function $\hat{p}_1(p_2)$ takes a piecewise linear form, as shown in Figure \ref{fig:PPPPE}. It strictly increases in $p_2$ when $p_2\in [0, H]$ and is flat when $p_2>H$. Moreover, if $\lambda^V>0$, then $p_2>H \Rightarrow \hat{p}_1(p_2) < \mathbb{E}[\omega]$. 
\end{lemma}

As previously established, the cutoff advance price, $\hat{p}_1(p_2)$, is derived from two intersecting constraints. The first constraint ensures that the plan is consistent with a PE, while the second ensures that it is consistent with a PPE when multiple PEs are present. Each of these constraints is determined as a function of $p_2$:

The first constraint, associated with PE, validates that \emph{pre-purchase} is locally optimal at $h=(p_1,p_2)\in T_1$, as per Condition \eqref{eq_def:PE}.

\begin{equation}
    p_1\leq 
    \begin{cases}
    p_2 &\textit{ if } p_2\leq L,\\
    (\lambda^V+1)L(1-q)L+q p_2  &\textit{ if } L < p_2 \leq (\lambda^V+1)L,\\
    \frac{(\lambda^V+1)(1-q)L+(\lambda^M+1)qp_2}{\lambda^M q+1}&\textit{ if } (\lambda^V+1)L\leq p_2 \leq H,\\
    \mathbb{E}[\omega] + \lambda^V q^2 H + \lambda^V (1-q^2)L &\textit{ if } p_2>H.
    \end{cases}\label{eq:pre-purchase_as_PE}
\end{equation}

The second constrain ensures that among multiple PEs, the consumer implements a plan that maximizes the expected utility, which is, therefore, consistent with a PPE. It imposes the following on the price $p_1$:

\begin{equation}
    p_1\leq 
    \begin{cases}
    p_2 &\textit{ if } p_2\leq L,\\
    (1-q)L+\lambda^V q(1-q)L+qp_2+\lambda^M(1-q)qp_2&\textit{ if } L < p_2 \leq H,\\
    \mathbb{E}[\omega]-\lambda^V(1-q)q(H-L) &\textit{ if } p_2>H.
    \end{cases}\label{eq:pre-purchase_as_locally_preferred}
\end{equation}

The second constraint is not necessarily stricter than the first one, as illustrated in Figure \ref{fig:PPPPE}, specifically when $p_2>\frac{(1-q)\lambda^V L}{q\lambda^M}$.

\paragraph{Interpreting Figure \ref{fig:PPPPE}.}\mbox{}\\

\vspace{-0.2in}

\par When $p_2\leq L$, $\hat{p}_1(p_2)=p_2$, indicating that it increases in $p_2$.
Consequently, the consumer's decision essentially becomes a choice between purchasing at the advance price $p_1$ at $T_1$ or waiting and buying at the spot price $p_2$ at $T_2$. The consumer will purchase at whichever price is lower.

When $L< p_2 \leq H$, $\hat{p}_1(p_2)$ continues increasing in $p_2$. 
This trend is driven by two distinct factors. 
First, aligns with the standard theory, the consumer's willingness to pay at $T_1$ grows as the share of she needs to pay after, which equals $q\cdot p_2$, increases. Second, the consumer faces anticipatory losses, which also increases in $p_2$. By choosing to pre-purchase, the consumer can mitigate such an anticipatory loss. The elaboration of this second mechanism is reserved for the next subsection.

When $p_2>H$, however, $\hat{p}_1(p_2)$ remains constant despite an increase in $p_2$ increases. 
In this range, a change in the spot price $p_2$ does not affect the $T_1$-consumer's behavior because the will never buy at $T_2$. Consequently, the consumer's decision effectively becomes a choice between purchasing at the advance price $p_1$ at $T_1$ or not purchasing at all. This simplifies the decision to a single-stage problem, where the consumer decides whether to accept a take-it-or-leave-it offer at price $p_1$. In such a scenario, choosing not to buy results in less fluctuation in the value dimension, making it more attractive to a loss-averse consumer. Therefore, to incentivize the consumer to accept this offer, the cutoff advance price must be set lower than $\mathbb{E}[\omega]$. This finding is consistent with the previous research indicating that loss aversion tends to reduce a consumer's willingness to buy under uncertainty (cf. \citealt{herweg2013newsvendor}).

The following corollary summarizes the immediately preceding paragraph:
\begin{corollary}\label{cor:too_high_p2_not_profitable}
For the seller, setting the spot price $p_2$ strictly greater than $H$ is not profitable, as this does not correspondingly raise the optimal advance price $\hat{p}_1(p_2)$, which peaks at $p_2=H$.
\end{corollary}

The following proposition articulates the first main result of this paper, establishing that the seller can charge an advance price exceeding $\mathbb{E}[\omega]$ when faced with a loss-averse consumer.

\begin{proposition}\label{prop:mainresult}
The optimal pricing $(p^{\ast}_1,p^{\ast}_2)$ is such that
	\begin{itemize}
	    \item $p^{\ast}_1=\min \{\frac{\mathbb{E}[\omega]+q\lambda^M H+ (1-q)\lambda^V L}{1+q\lambda^M}, \mathbb{E}[\omega]+\lambda^V q (1-q)L+\lambda^M q (1-q)H\}$;
	    \item $p^{\ast}_1$ is strictly greater than $\mathbb{E}[\omega]$ if either of $\lambda^V$ or $\lambda^M$ is positive;
	    \item $p^{\ast}_2=H$.
	\end{itemize}

\end{proposition}

The optimal advance price $p^{\ast}_1$ needs to satisfy both constraints of \eqref{eq:pre-purchase_as_PE} and \eqref{eq:pre-purchase_as_locally_preferred}. Both of these constraints are greater than $\mathbb{E}[\omega]$ at $p_2=H$.
Given that $p^{\ast}_1$ is also the seller's profit, the result indicates that the seller can extract profit from the consumer's loss aversion, yielding a profit greater than $\mathbb{E}[\omega]$, the maximum profit expected when encountering a consumer with a standard risk-neutral preference.

Proposition \ref{prop:mainresult} poses an intriguing paradox in the behavior of a loss-averse consumer. Suppose the consumer adopts a plan to never make a purchase in either stage given the prices. In that case, her $T_1$ expected utility, given that the corresponding reference distribution is constituted by this plan, has the value of $0$. However, in the Preferred Personal Equilibrium (PPE), where the $T_1$-consumer chooses to pre-purchase, the expected utility becomes negative, which is a value lower than that under the never-purchase plan!\footnote{It turns out that the main insight--loss aversion increases the worker's willingness to pre-purchase in this dynamic decision problem--remains robust under the assumption of a lagged-belief reference distribution, as per \cite{KR09}, provided that the parameters $\lambda^M$ and $\lambda^V$ are not too small. Thus, the main finding does not critically rely on Assumption \ref{assu:recent_belief} and is not overturned even with a strict application of the \cite{KR09} model. Further details are available upon request.}

One might question why the never-purchase plan is not a preferred personal equilibrium at the optimal prices $(p^{\ast}_1,p^{\ast}_2)$. The explanation is rooted in Assumption \ref{assu:recent_belief}, which states that the $T_2$-consumer has her own reference distribution (re-)constructed according to her recent beliefs. This prevents the $T_1$-consumer from ruling the $T_2$-consumer sub-plan. Specifically, when the state is $\omega=H$ and the $T_2$-consumer is presented with $p_2^{\ast}=H$, she will be guided by her locally favorable sub-plan to choose to buy, even though this action may be less desirable for the $T_1$-consumer. As a result, despite being a personal equilibrium (PE) that guarantees a higher expected utility for the $T_1$-consumer, the never-purchase plan fails to be in a PPE.

\begin{corollary}\label{cor:SOSDed}
Given that $p_2=H$, the $T_1$-consumer has a strict incentive to pre-purchase at $p_1=\mathbb{E}[\omega]$.
\end{corollary}

\paragraph{What dimension of losses increases the cutoff advance price?}

\begin{corollary}\label{cor:monotone_money_commit}
Given $p_2=H$, the consumer's monetary-dimension loss aversion increases her willingness to pre-purchase.
\end{corollary}

If $\lambda^M>0$ and $\lambda^V=0$, the condition $\hat{p}_1(H)>\mathbb{E}[\omega]$ is satisfied. This suggests that at $T_1$, a loss-averse consumer, compared to one with standard preference, overreacts to a future increase in price.

\begin{corollary}\label{cor:monotone_value_commit}
	Given $p_2=H$, the consumer's value-dimension loss aversion increases her willingness to pre-purchase.
\end{corollary}

If $\lambda^M=0$ and $\lambda^V >0$, $\hat{p}_1(H) >\mathbb{E}[\omega]$ also holds.
For a price $p_2=H$, the $T_2$-consumer will choose to reject in the state $\omega=L$. Given that $L>0$, consuming nothing, which is the expected consequence if she chooses to wait and $\omega=L$, is less appealing than consuming the good.
As a result, if the $T_1$-consumer chooses to wait, she anticipates a greater loss in the value dimension.
Therefore, the $T_1$-consumer's aversion to the potential value-dimension loss increases the appeal of pre-purchasing.

The observation in Corollary \ref{cor:monotone_value_commit} parallels the ``attachment effect'' as described by \cite{kHoszegi2006model}, but operates differently from previous literature. Some scholars characterize the ``attachment effect'' as the spanning-across-periods influence of prior expectations about acquiring a good on future consumption behavior, and emphasize the role of aversion to contemporary value-dimension loss, which stems from the received consumption value failing to meet the expected one.\footnote{
For example, \cite{rosato2016selling} demonstrated that introducing uncertainty about the availability of a low-priced substitute (a loss leader) can increase consumer's willingness to buy. This effect relies heavily on the assumption that consumers' reference distributions are formed before the uncertainty is resolved.
\label{footnote:exist_attachment_effect}}
In contrast, Corollary \ref{cor:monotone_value_commit} derives from the anticipatory (or prospective) loss in the value dimension, which arises from the perception that the expected consumption value may fall short of the current expectation. Thus, the underlying mechanism of this corollary could be seen as a form of ``prospective attachment effect''.

Drawing from Corollaries \ref{cor:monotone_money_commit} and \ref{cor:monotone_value_commit}, both monetary-dimension and value-dimension loss aversion unambiguously increase the $T_1$-consumer's inclination to pre-purchase.

\section{Seller's Price Commitment vs Flexibility}\label{sec:commitment}

The existing literature on the advance purchasing problem often assumes the seller commits to the spot price prior to the advance-purchase stage.\footnote{For example, \cite{gale1992airticket,gale1993advance,dana1998advance,shugan2000advance,prasad2011advance,karle2020selling}} This section enriches this literature by offering a fresh justification for this assumption. Specifically, I allow the seller to decide in advance whether to commit to the spot price.\footnote{An exception is the work of \cite{xie2001advancesell}, where they propose a scenario where the seller announces the spot price in advance. In their model, the credibility of this announcement is maintained by high marginal production costs. Their work focuses on which factor can maintain the credibility, while this paper focuses on whether the seller has an incentive to commit if this is not omitted.}

In cases where the seller gains no further information beyond the consumer's pre-purchase choice, there is no incentive to adjust the spot price subsequently. Therefore, it is more interesting to focus on the contrasting case where the seller has perfect information:

\begin{assumption}\label{assumption:perfectlearning}
The seller will learn the state of nature $\omega$ at the beginning of the spot-market stage.
\end{assumption}

This assumption applies to industries that allow last-minute price changes, or markets in which monopolists obtain big data that allows them to learn consumers' demands. In the general case, the assumption provides a useful benchmark because it maximizes the informational gain the seller can obtain by choosing not to commit to the spot price.

\subsection{Benchmark: A Consumer with Risk-neutral Preference}
I will first discuss the situation in which the consumer exhibits a standard preference, namely, $\lambda^M=\lambda^V=0$.

\begin{proposition}\label{prop_sec_4.1}
The seller is indifferent between making a spot-price commitment or not if the consumer has a standard risk-neutral preference.
\end{proposition}

However, this indifference vanishes with any degree of consumer risk aversion. Commitment to the spot price is never optimal for the seller if the consumer's utility function is concave utility function with respect to material payoffs, reflecting risk aversion. In essence, under standard preference, offering an advance-purchase option is not profitable. This observation, while briefly introduced here, is explored in detail in Section \ref{sec:RA}.

\subsection{A Consumer with Belief-based Loss-averse Preference}

Now consider scenarios where at least one of $\lambda^V$ and $\lambda^M$ is positive. The following observation can be derived directly from Lemma \ref{lemma:2ndstage}.

\begin{remark}
Conditional on the seller not committing to the spot price, it is optimal for him to adjust the spot price, $p^{\omega}_2$, to be equal to $\omega\in \{H,L\}$, where $p_2^{\omega}$ represents the adjusted price in that state.
\end{remark}

As Section \ref{sec:results} has already covered the scenario in which the seller has committed to the spot price, the following discussion will focus on the contrasting case where the seller has not committed to the spot price.

\begin{lemma}\label{prop:optpricingnocommit}
Conditional on the seller not committing to the spot price, the optimal pricing is $(p_1,p_2^H,p_2^L)=(\frac{\mathbb{E}[\omega]+q\lambda^M H}{1+q\lambda^M},H,L)$. With this pricing, the seller obtains a profit $\pi_{nc}$ strictly greater than $\mathbb{E}[\omega]$ by selling the good in the advance-purchase stage.
\end{lemma}

This lemma suggests that, with price flexibility, offering advance sales can be more profitable than in scenarios without advance sales. In this case, the consumer's aversion to monetary-dimension loss increases her willingness to pre-purchase, whereas her aversion to value-dimension losses has no impact.

\begin{proposition}\label{prop: to_commit}
If $\lambda^V>0$, a seller who can perfectly learn the consumer's demand after $T_1$ strictly prefers to make a spot-price commitment in advance.
\end{proposition}

Choosing not to commit to the spot price results in a lower profit when contrasted with the maximum profit a seller could make, as stated in Proposition \ref{prop:mainresult}. This intriguing result implies that allowing a perfect price adjustment backfires, which is somewhat counter-intuitive.

This counter-intuitive finding provides an explanation for the causally observed practice of sellers maintaining a high ``regular price'' in the spot market, even though they are not obligated to do so.

\paragraph{Why committing to a spot-price yields higher profits?}
\mbox{}\\

An initially plausible but incorrect interpretation might suggest that the consumer's aversion to monetary-dimension loss, assumed to be higher under a commitment, drives consumer behavior. This line of thought posits that the seller's commitment to a high spot price, precluding price reductions, could lead to significant monetary loss for a consumer who chooses to wait, as the spot price will definitely exceed the advance price.  Without this commitment, however, the seller's flexibility to adjust the spot price downward in state $\omega=L$ mitigates this loss, making the option to wait more attractive. However, this view incorrectly assumes that the spot-price commitment results in a higher expenditure in $T_2$. Lemma \ref{lemma:2ndstage} contradicts this by indicating that the $T_2$-consumer would reject a spot price exceeding $L$ if $\omega=L$. Consequently, if she chooses to wait, the consumer's monetary-dimension loss is not necessarily higher under the seller's spot-price commitment, and thus does not explain the increased willingness to pre-purchase under such a scenario.

The correct interpretation hinges on the consumer's aversion to value-dimension loss: under a high spot-price commitment, choosing to wait incurs the risk of not consuming, resulting in $0$ consumption utility when the state is $\omega=L$. This outcome could be more disappointing than if she chooses to pre-purchase, which ensures that her consumption utility in this state would be $L$. Consequently, the consumer's aversion to loss in the value dimension makes her willing to accept a higher advance price when this commitment exists. I refer to this as the ``prospective attachment effect'' to distinguish it from the existing discussion of attachment effect. However, this effect disappears in the absence of a seller's price commitment. If the seller chooses not to commit to the spot price, thus having the flexibility to adjust the price, he can always sell the good to the consumer at an adjusted price. As a result, the consumer is relieved of the fear of ending up without the good.

\section{Further Analysis}\label{sec:further}

\subsection{A Consumer with Standard Risk-averse Preference}\label{sec:RA}

This section presents the observations for a scenario where the consumer is risk-averse, who maximizes a strictly concave function of material payoffs. These observations contrast with the main analysis involving a loss-averse consumer, thereby highlighting the differences between standard risk aversion and loss aversion within the context of an advance-purchase game.

\begin{proposition}\label{prop:riskaverse}
If the consumer is risk-averse, the following holds:
\begin{itemize}
    \item[(i)] The consumer's cutoff advance price is strictly less than $\mathbb{E}[\omega]$.
    
    \item[(ii)] 
    A seller, expected to learn the state of nature perfectly in the second stage, strictly prefers not to commit to the spot price in advance

\end{itemize}
\end{proposition}

Statement (i)  implies that risk aversion reduces the consumer's willingness to pre-purchase, distinguishing it from the impact of loss aversion.
Specifically, in the standard framework, which assumes no narrow bracketing, a consumer who chooses to pre-purchase is exposed to the uncertainty of the consumption value. By choosing to wait, however, the consumer gains the flexibility to make an informed decision once the value is realized, thus avoiding this risk.  Consequently, the consumer is inclined to wait unless the seller offers a discount on the advance price. This risk discount, lowering the advance price below $\mathbb{E}[\omega]$, induces the consumer to pre-purchase.

Statement (ii) suggests that when encountering a risk-averse consumer, the spot-price commitment reduces the seller's profitability.\footnote{In this case, the seller can still achieve a maximum profit of $\mathbb{E}[\omega]$. However, it is conditional on the avoidance of a prior commitment at $p_2$, coupled with the guaranteed sale of the good at $T_2$.}
This also contrasts with the statement made in  Proposition \ref{prop: to_commit}.

\subsection{Advance Purchasing under an Initial-Belief Reference Distribution}\label{sec:fast_play}

In the previous sections, it was shown that under Assumption \ref{assu:recent_belief} that a consumer's reference distribution is shaped by recent beliefs, loss aversion increases the inclination to pre-purchase.

This subsection aims to shed light on how the absence of influence from the previous stage, a consequence of Assumption  \ref{assu:recent_belief}, derives the consumer behavior. To do this, I examine and contrast the consumer behavior under Assumption \ref{assu:recent_belief} with the behavior observed under an alternative assumption where the reference distribution remains static and does not adjust to recent beliefs when transitioning to the second stage.

Hereafter, consider a scenario under an alternative assumption:

\begin{assumption}\label{assu:initial_belief}
In each decision stage, the consumer's reference distribution is determined by her initial belief, which is her belief established at the beginning of the advance-purchase stage.
\end{assumption}

As Lemma \ref{lemma:pref_agree_under_same_reference} will illustrate, this assumption, widely adopted in applied research, serves as a basis for applying \cite{kHoszegi2006model}'s one-stage model to conduct a ``static analysis'' in dynamic decision-making problems.\footnote{Owing to its parsimony, the static analysis has been widely adopted, including works by \cite{herweg2013newsvendor,heidhues2014sales,herweg2015loss,karle2020selling,herweg2010binary,herweg2013flatrate}.\label{ft:static-analysis_in_lit}}\footnote{A relevant example is the work of \cite{karle2020selling}, who similarly explore the role of loss aversion in advance-purchasing problems, with particular emphasis on the impact of consumer information in the early stage on market performance under monopolistic and duopolistic conditions.}

Under Assumption \ref{assu:initial_belief}, the $T_2$-consumer's reference distribution is inherited from the $T_1$-consumer and remains unchanged, even after the value of $\omega$ is realized. Thus, it is shaped by the $T_1$-consumer's plan, as the plan forms part of her initial beliefs. This is mathematically represented in the consumer's expected utility for choosing an action, $a_c\in A_c(h)$:

\begin{equation}
    \begin{array}{lll}
        \Tilde{u}_{c,h}(a_c,\alpha_c(\cdot|\Bar{h}))
        &:= \mathbb{E}_{a_c} \left\{ \sum \limits_{\tilde{z}\in Z(h)} \alpha_c(\tilde{z}|\Bar{h}) \cdot u_c(g_c(\cdot),g_c(\tilde{z})) \bigg|h,a_c
        \right\}\\
         &= \sum \limits_{z\in Z(h,a_c)}{\alpha_c(z|h,a_c)} \cdot \left\{ \sum \limits_{\tilde{z}\in Z(h)}{\alpha_c(\tilde{z}|\Bar{h})}\cdot \big[ u_c\big(g_c(z), g_c(\tilde{z})\big) \big]  \right\} .
    \end{array}
\end{equation}

The property that the consumer's current-stage reference distribution is constructed by her conditional belief at the advance-purchase stage now is reflected by the second argument of this function, $\alpha_c(\cdot|\Bar{h})$, rather than the recent conditional belief $\alpha_c(\cdot|h)$. Here, the node $\Bar{h}$ denotes the corresponding node in $T_1$ that satisfies (i) $\Bar{h}=h$ if the on-focus node $h$ is in $T_1$ and (ii) $\Bar{h}\in T_1: \Bar{h}\prec h$ if $h\in T_2$. For instance, given a node $h=((p'_1,p'_2), wait, H)$ in $T_2$, the corresponding reference distribution is determined by the conditional belief $\alpha_c(\cdot|\Bar{h})$ at $\Bar{h}=(p'_1,p'_2)$. This function coincides with \eqref{eq:U_function} at $h\in T_1$, and differs from it at $h\in T_2$. 
Given this expected utility, the $T_2$-consumer would suffer losses if her sub-plan diverges from the $T_1$-consumer's plan, which determines the $T_2$ reference distribution.

The following lemma formally establishes that Assumption \ref{assu:initial_belief} allows a ``static analysis'' on dynamic decision problems. For convenience, the expected utility at $h$  is denoted by $\mathbb{E}_{\alpha_c}\left[\tilde{u}_c|h\right] := \sum_{a'_c\in A_c(h)}\alpha_c(a'_c|h)\cdot \Tilde{u}_{c,h}(a_c,\alpha'_c(\cdot|\Bar{h}))$.

\begin{lemma}
    For any $\Bar{h}\in T_1$ and $h'\in T_2$ s.t. $h'=(\Bar{h},wait,\omega)$, it holds true that the expected utility by choosing $a_c\in A_c(h')$ satisfies that
    \begin{equation}
    \begin{array}{ll}         
         \Tilde{u}_{c,\Bar{h}} (a_c,\alpha_c(\cdot|\Bar{h}))& = \mathbb{E}_{\alpha_c}\left[ \mathbb{E}_{\alpha_c}\left[\tilde{u}_c|h'\right]   \bigg|\Bar{h},a_c
        \right] \\
        &=q \cdot \mathbb{E}_{\alpha_c}\left[\tilde{u}_c|(\Bar{h},wait,H) \right]  + (1-q)  \cdot \mathbb{E}_{\alpha_c}\left[\tilde{u}_c|(\Bar{h},wait,L)\right]
    \end{array}
    \end{equation}\label{lemma:pref_agree_under_same_reference}
\end{lemma}

This lemma underscores that the $T_1$-consumer's expected utility is essentially her expectation of the potential expected utilities of the $T_2$-consumer. Consequently, a plan that maximizes the $T_1$-consumer's expected utility also optimizes the $T_2$-consumer's. 
This property holds across different application contexts when Assumption \ref{assu:initial_belief} is in place.

The Preferred Personal Equilibrium (PPE), as outlined in Definition \ref{defn:PPE}, continues to serve as the solution concept for analyzing consumer behavior in this section. It is important to note, however, that under Assumption \ref{assu:initial_belief}, the expected utilities in conditions (i) and (ii) of the definition undergo a modification to $\Tilde{u}(\cdot,\alpha_c(\cdot|\Bar{h})$.

\begin{remark}\label{remark:PPE_in_static}
    Under Assumption \ref{assu:initial_belief}, a plan consistent with a Preferred Personal Equilibrium (PPE) is the the one that maximizes the $T_1$-consumer's expected utility among all plans consistent with a Personal Equilibrium (PE). 
\end{remark}

The concept of the Preferred Personal Equilibrium (PPE) coincides with the original static PPE as proposed by \cite{kHoszegi2006model}.

Let $ \bar{p}_2(\lambda^V,\lambda^M):=\frac{\mathbb{E}[\omega]-(1-q)\lambda^Vq(H-L)}{1+\lambda^M}$, a price threshold that falls strictly below $p_2=H$ whenever either $\lambda^V$ or $\lambda^M$ is positive.

\begin{proposition}\label{prop:fast_play_pre-purchasing}
    Under Assumption \ref{assu:initial_belief}, for $p_2\geq \bar{p}_2(\lambda^V,\lambda^M)$, a loss-averse consumer following a PPE will have her cutoff advance price not exceed $\mathbb{E}[\omega]$.
\end{proposition}

Under Assumption \ref{assu:initial_belief}, if $p_2$ exceeds $\bar{p}_2(\lambda^V,\lambda^M)$, the seller cannot sell the good in advance at a price beyond its expected value $\mathbb{E}[\omega]$, in stark contrast to Proposition \ref{prop:mainresult}. This divergence underscores how the (non-)updating of the reference distribution can yield qualitatively opposite predictions in similar contexts.
To illustrate, consider the case where $(p_1,p_2)=(\mathbb{E}[\omega], H)$. In this case, a never-purchase plan, in which the consumer chooses not to buy in either stage, is validated as a personal equilibrium, and it guarantees the consumer an expected utility of zero.
Conversely, any plan that assigns a positive chance to buy in any stage induces an anticipatory value-dimension loss derived from the fluctuation of the consumption value. This occurs, for example, when the consumer receives $L$ or $0$ consumption utility and compares it to a reference distribution that allocates a positive probability to $r^V=H$. Consequently, a loss-averse consumer who adheres to an initial-belief reference distribution is inclined toward the never-purchase plan to avoid these losses.
Therefore, to engage such a consumer to but when $\lambda^V>0$, the seller must offer a positive rent, reducing his expected profit to less than $\mathbb{E}[\omega]$. This mechanism reflects how value-dimension loss aversion reduces the consumer's willingness to pre-purchase.\footnote{This observation echoes the Proposition 1 of \cite{karle2020selling}}.

\section{Concluding Remarks}
\label{sec:conclusion}

The analysis of the advance-purchase game suggests that loss aversion can increase a consumer's willingness to pre-purchase, thereby boosting the monopolist's profit. Specifically, loss aversion in both monetary and value dimensions can lead the consumer to pre-purchase at a price that exceeds the expected value of the good---the maximum price that a standard risk-neutral consumer is willing to pay for the good. On the one hand, the monetary-dimension loss is more severe for a consumer who chooses to wait, as she might have to pay a price $p_2$ exceeding $p_1$ in case of state $\omega=H$. On the other hand, the value-dimension loss is also more severe for this consumer in the case of state $\omega=L$, as she could end up with no consumption, which is associated with a consumption utility of $0$. Thus, both losses make the option of waiting less attractive for the $T_1$-consumer. Furthermore, this observation implies that the effect of loss aversion may qualitatively differ from that of standard risk aversion.

Moreover, the study also reveals that a monopolist's ability to exploit the consumer's loss aversion is subject to constraints. As shown in Corollary \ref{cor:too_high_p2_not_profitable}, an excessively high spot price reduces the consumer's acceptable advance price, causing it to fall below the expected value of the good. This observation mirrors real-world intuition, as seen when a consumer encounters a promotional item presented with both a current discounted price and a (non-binding) manufacturer's suggested retail price (MSRP). An unrealistically high MSRP may lose its effectiveness in getting the consumer to buy it now.

My second finding offers a fresh interpretation of a prevalent assumption in the advance-purchase literature that the seller commits to the spot price when offering an advance sale. Though not obligatory, my analysis reveals that the seller, when faced with a loss-averse consumer, would still choose to make this commitment. Without it, the seller retains the flexibility to adjust the spot price in response to market information and can still preserve the profitability of an advance option compared to not offering it. However, this flexibility restricts the seller from exploiting the $T_1$-consumer's aversion to value-dimension loss, and consequently reduces the seller's profit. Recognizing this limitation, the seller is thus incentivized to commit in advance to a high-level spot price, thereby exploiting the consumer's aversion to value-dimension loss to enhance profitability.

The adoption of the recent-belief reference distribution, outlined in Assumption \ref{assu:recent_belief}, diverges from the model by \cite{KR09}, which bases the reference distribution on the lagged belief. However, some empirical work such as \cite{smith2019lagged} and \cite{song2016experiment_on_ref}, supports this assumption by showing that references quickly adjust to new information. This assumption is particularly relevant in scenarios with ample time between learning the state of nature and the subsequent decision stage, allowing the consumer to update her reference distribution according to the most recent belief.

To further explore the role of dynamic updating of the reference distribution, Section \ref{sec:fast_play} examines the alternative scenario where the reference distribution remains static once established, similar to \cite{karle2020selling}.\footnote{Their focus may seem more plausible when the time interval between pre-sale and consumption is small.} I observe a reversal of the loss aversion effect, resulting in a decreased willingness to pre-purchase. Thus, it demonstrates that the update of the reference distribution matters on the effect of loss aversion, suggesting the need for further empirical exploration.

These contrasting observations not only advance our understanding of loss aversion in similar dynamic settings but also suggest intriguing speculation. If this difference is noticed by a seller, he may want to strategically extend the time interval between the two decision stages to increase the profitability of advance sales. This, in turn, would shape certain business practices, such as avoiding advance sales in a short time before consumers will recognize their needs, or temporarily removing the product from the market after an advance sale.

\clearpage
\begin{doublespacing}   
\bibliographystyle{chicago}
\bibliography{bib}

\end{doublespacing}
\clearpage

\pagenumbering{roman}
\appendix

\section{The opportunity to trade in the spot market}\label{sec:2nd_stage}

In this section, I explore an alternative scenario in which the consumer's decision-making process is limited to the advance-purchase stage, thus excluding her spot-market stage buying option. The consumer thus faces a one-stage decision problem with uncertainty about the state of nature. The analysis aims to shed light on how the absence of a second purchase opportunity influences the effect of loss aversion.

As suggested by Proposition \ref{prop:mainresult}, the seller and the consumer will trade in the advance-purchase stage rather than in the spot-market stage. Nevertheless, the mere presence of the spot-market stage, though not actively participated in, significantly impacts the role of consumer loss aversion. The following analysis will show that, with the absence of the spot-stage buying option, the effect of loss aversion is reversed---it reduces the consumer's willingness to buy and thus reduces the seller’s profit. 

\begin{remark}\label{remark_no2ndchance}
    Suppose the spot-market trade is not available. The highest acceptable advance price for the consumer becomes $\mathbb{E}[\omega]-(1-q)\lambda^V\cdot q( H-L) $, which is strictly lower than $\mathbb{E}[\omega]$ as long as $\lambda^V>0$.
\end{remark}

This remark follows directly from Lemma \ref{lemma:monotone} when considering the case of $p_2>H$. This follows from the fact that in the original game, the $T_2$-consumer would never choose to buy at such an high spot price.

When considering a consumer with standard risk-neutral preference (i.e., $\lambda^V=\lambda^M=0$), the seller's profit is $\mathbb{E}[\omega]$, identical to what he would receive if spot-market trading were feasible.

Shifting the focus to a consumer with aversion to value-dimension loss ($\lambda^V>0$), the consumer's loss aversion leads to a reduction in the consumer's willingness to buy. This result is due to the fluctuation of the consumption value when the consumer chooses to purchase, while abstaining from purchase guarantees a deterministic consumption value of $0$. 
This finding is comparable to Herweg's conclusions in a newsvendor problem, where loss aversion was found to suppress willingness to buy \citep{herweg2013newsvendor}.\footnote{Note that this scenario differs from Herweg's in two key respects: (i) he employs the choice-acclimating personal equilibrium (CPE) solution, and (ii) he assumes broad bracketing, with the newsvendor measuring losses in one-dimensional profits.}
Despite these consistencies, note that as shown in Corollary \ref{cor:SOSDed}, such conclusions do not always extend to multi-stage decision problems.

\section{Proofs}\label{sec:appen_proofs}

\begin{proof}[Proof of Proposition \ref{prop:standard}]

Suppose the seller commits to $p_2$. Then the seller can at most obtain expected profit $\mathbb{E}[\omega]$. 

In the case that the consumer waits at $T_1$:

\begin{itemize}
\item If $p_2\leq L$, the $T_2$-consumer will purchase, and expects to get $\mathbb{E}[\omega]-p_2$.

\item If $p_2\in (L, H]$, the $T_2$-consumer will purchase whenever $\omega=H$ and expects to get $q \cdot (H-p_2)$.

\item If $p_2>H$, the $T_2$-consumer will not purchase and expects to get $0$.

\end{itemize} 

To make the consumer to pre-purchase, the seller sets $p_1=\mathbb{E}[\omega]$. This maximizes the seller's expected profit: Suppose there is another pricing that gives the seller a strictly higher profit. The consumer expects a negative material payoff and she will deviate to never purchase in either stage.

Suppose the seller does not commit, he can still obtain $\mathbb{E}[\omega]$ by making the consumer pre-purchase at $T_1$. 
At $T_2$ the seller will set $p_2^\omega=\omega$. The consumer who anticipates this expects to get $0$ if she chooses to wait at $T_1$. Therefore, she will pre-purchase whenever $p_1\leq \mathbb{E}[\omega]$. In this case, the seller can obtain $\mathbb{E}[\omega]$ by selling the good at $T_1$.

\end{proof}

\begin{proof}[Proof of Lemma \ref{lemma:2ndstage}]

Suppose the $T_1$-consumer did not pre-purchase, the $T_2$-consumer chooses to buy or not after realizing the state $\omega$. Let $\bar{p}$ be her cutoff price below which she would buy.

		\begin{enumerate}
			\item[(i)] 
			Given $p_2> \bar{p}$, if the $T_2$-consumer is expected to not to purchase. Her utility is $0-\lambda^V \cdot 0 =0$ by carrying out such an expected plan.
			If she instead expects to purchase, and carry out the plan, her utility is $\omega-p_2-\lambda^M(p_2-0)$.
			Conditional on that $\bar{p}$ satisfies $\bar{p}> \frac{\omega}{1+\lambda^M}$, the $T_2$-consumer's credible plan at that node prescribes her not to buy the good.

			\item[(ii)] 
			Given $p_2\leq \bar{p}$, if the $T_2$-consumer expects to purchase. Her utility is $\omega-p_2-\lambda^M(\max\{0, p_2-\bar{p}\})=\omega-p_2$ by carrying out this plan.
		    If she instead expects not to purchase, by carrying out this plan, her utility is $0-\lambda^V\cdot \omega$. 
			Conditional on that $\bar{p}$ satisfies $\bar{p}\leq (1+\lambda^V) \omega$, the $T_2$-consumer's credible plan at that node prescribes her to buy the good.

		\end{enumerate}
		
		In sum, (a) given a $\omega$, a cutoff price at $T_2$ within the range of $[\frac{\omega}{1+\lambda^M}, (1+\lambda^V)\omega]$ can be supported by some credible plan at that node. (b) within this range, the level of $p_2$ that makes the consumer be indifferent between carrying out either the purchasing plan or the non-purchasing plan is $\omega$. Thus, $p_2=\omega$ is the cutoff price in PPE.

\end{proof}

Lemma \ref{lemma:monotone} is immediately from Lemma \ref{lemma:LPPE}.

   The following lemma concludes the behavior in the PPE
        
        \begin{lemma}[PPE behavior]\label{lemma:LPPE} If the player adopts the PPE, then:
            \begin{itemize}
                \item Given that $p_2\leq L$, 
     the consumer pre-purchases whenever $p_1\leq p_2$, otherwise, she will purchase at $T_2$ whenever $p_1> p_2$.

                \item Given that $L<p_2<(\lambda^V+1)L$, the consumer pre-purchases whenever
                \begin{equation*}
                    p_1\leq \min \{(1-q)L+\lambda^V q (1-q)L+q p_2+\lambda^M(1-q)q p_2, (\lambda^V+1)(1-q)L+q p_2 \},
                \end{equation*}
                otherwise, she waits at $T_1$ and then purchases at $T_2$ whenever $\omega=H$

                \item Given that $(\lambda^V+1)L<p_2<H$, the consumer pre-purchases whenever
              \begin{equation*}
                    p_1\leq \min \{(1-q)L+\lambda^V q (1-q)L+q p_2+\lambda^M(1-q)q p_2,\frac{(\lambda^V+1)(1-q)L+q(\lambda^M+1)p_2}{1+q\lambda^M}\},
                \end{equation*}
                otherwise, she waits at $T_1$ and then purchases at $T_2$ whenever $\omega=H$.
                
                \item Given that $p_2>H$, the consumer pre-purchases whenever
                \[p_1\leq \mathbb{E}[\omega]-\lambda^V (1-q)q(H-L),\] otherwise, she will not purchase in either stage.

            \end{itemize}
            
        \end{lemma}
    
\begin{proof}

I first figure out the behavior of the $T_1$-consumer that is supported by a personal equilibrium (PE). Recall that a plan is consistent with a PE if any deviation from it yields a weakly lower expected utility.
 
As per Section \ref{sec:main_loss_averse_consumer_pre-purchasing}, there are three sub-cases that need to be discussed.

 \paragraph{Sub-case $p_2\leq L$:}

Choosing to pre-purchase is consistent with the personal equilibrium whenever the expected utility specified in \eqref{eq:implement_plan_pp} exceeds that of deviating to wait, detailed in \eqref{eq:p2_low_pp_wait}.  
\begin{equation*}
    qH + (1-q)L -p_1 -(1-q)\cdot q \cdot \lambda^V(H-L) \geq qH + (1-q)L -p_2 -(1-q)\cdot q \cdot \lambda^V (H-L) - \lambda^M \mathcal{L} (p_2-p_1).
\end{equation*}

Choosing to wait is consistent with the personal equilibrium whenever the expected utility specified in \eqref{eq:implement_plan_wait} exceeds that of deviating to wait, detailed in \eqref{eq:deviate_plan_wait_to_pp}. 
\begin{equation*}
    q H+(1-q)L-p_2-(1-q)q\lambda^V(H-L) \geq q\cdot H+(1-q)L-p_1-(1-q)q\lambda^V(H-L) -\lambda^M\mathcal{L}(p_1-p_2).
\end{equation*}

 \paragraph{Sub-case $L<p_2\leq H$:}
Choosing to pre-purchase is consistent with the personal equilibrium if the following condition holds: 
\begin{equation*}
     qH + (1-q)L -p_1 -(1-q)\cdot q \cdot \lambda^V(H-L) \geq q\cdot \big\{H - p_2 - \lambda^M\mathcal{L}(p_2-p_1) \big\} + (1-q)\cdot \big\{-q\lambda^V\cdot H - (1-q)\lambda^V\cdot L \big \}.
\end{equation*}

Choosing to wait is consistent with the personal equilibrium if the following condition holds: 
\begin{align*}
    &q\cdot \big\{ H-p_2-(1-q)\lambda^M \cdot p_2\} + (1-q) \cdot \{q \cdot [0-\lambda^V\cdot H] \big\}\\
    &\geq q\cdot H +(1-q)L-p_1 - (1-q)q\lambda^V (H-L)-q\lambda^M\mathcal{L}(p_1-p_2)-(1-q)\lambda^M\cdot p_1.
\end{align*}

 \paragraph{Sub-case $p_2>H$:}
 
Choosing to pre-purchase is consistent with the personal equilibrium if the following condition holds: 
\begin{equation*}
     qH + (1-q)L -p_1 -(1-q)\cdot q \cdot \lambda^V(H-L) \geq 0 -q \cdot \lambda^V \cdot H -(1-q) \cdot \lambda^V \cdot L .
\end{equation*}

Choosing to wait is consistent with the personal equilibrium if the following condition holds: 
\begin{equation*}
    0 \geq  q \cdot H + (1-q) \cdot L -\lambda^M\cdot  p_1.
\end{equation*}

In summary, the $T_1$-consumer following a PE has behavior characterized as follows. 

Furthermore, this I compare expected utilities across different PEs when multiple PEs exist. A PE that is either unique or yields the highest utility is identified as a Preferred Personal Equilibrium (PPE).

\begin{itemize}
	\item $p_2\leq L \ \&\  p_1>p_2$
	\emph{to wait at $T_1$ and buy it regardless of the state at $T_2$} is the only PE at $T_1$.

	\item $p_2\leq L \ \&\  p_1\leq p_2$
	
	only \emph{pre-purchasing} is supported by a PE.

    \item $L <p_2\leq H \ \&\  p_1>p_2$

	\begin{itemize}
	    \item when $p_2\leq (\lambda^V+1)L$, \emph{pre-purchasing} is supported to be a PE if $p_1 \leq (\lambda^V+1) (1-q) L +q\cdot p_2$, whereas when $p_2>(\lambda^V+1)L$, \emph{pre-purchasing} is not a PE.
    
        \item \emph{waiting} is supported to be a PE if $p_1\geq \frac{q p_2 +\lambda^M(2q-q^2)p_2+(1-q)L+\lambda^V (1-q)qL}{1+\lambda^M}$. Moreover, it will always be true for $p_1>p_2$ if $p_2>\frac{L+\lambda^V q L}{1+\lambda^M(1-q)} $.

	    \item (existence) If $p_2<  \frac{\lambda^V (1-q)L+\lambda^M \lambda^V L+\lambda^M L}{\lambda^M q}$,
	    
	    $\left(\frac{q p_2 +\lambda^M(2q-q^2)p_2+(1-q)L+\lambda^V (1-q)qL}{1+\lambda^M} , (\lambda^V+1) (1-q) L +q\cdot p_2\right)$ is nonempty. 
        
        On the other hand, if $p_2> \frac{\lambda^V (1-q)L+\lambda^M \lambda^V L+\lambda^M L}{\lambda^M q}$ implies that waiting is always a PE, for $\frac{\lambda^V (1-q)L+\lambda^M \lambda^V L+\lambda^M L}{\lambda^M q}>\frac{L+\lambda^V q L}{1+\lambda^M (1-q)}$.

        \item If $p_1\leq (1-q)L+\lambda^V (1-q)qL +q p_2+\lambda^M(1-q)q p_2 $, to pre-purchase, as a PE, gives a higher utility than the PE assigning the consumer to wait.
        
        \item In sum, in the PPE, when
        $\min\{(1-q)L+\lambda^V (1-q)qL +q p_2+\lambda^M(1-q)q p_2,  (\lambda^V+1) (1-q) L +q\cdot p_2\} \geq p_2$, then the consumer will pre-purchase when $p_1>p_2$ and
        \begin{equation*}
             p_1 < \min\{(1-q)L+\lambda^V (1-q)qL +q p_2+\lambda^M(1-q)q p_2,  (\lambda^V+1) (1-q) L +q\cdot p_2\}. 
        \end{equation*}
        
       	\end{itemize}

	\item $L<p_2\leq H\  \&\  p_1\leq p_2$
	
	\begin{itemize}
	    \item \emph{pre-purchasing} is consistent with a PE if $p_1\leq \frac{(\lambda^V+1)(1-q)L+q(\lambda^M+1)p_2}{1+q\lambda^M}$

	    Moreover, if $p_2<(\lambda^V+1)L$, $p_1\leq p_2$ satisfies this condition, meaning pre-purchasing is always a PE.
	    
	    \item \emph{waiting} is a PE if $p_1\geq \frac{q p_2+\lambda^M (1-q)qp_2+(1-q)L+\lambda^V (1-q)q L}{1+\lambda^M (1-q)}$. Moreover, if $p_2< \frac{L+\lambda^V q L}{1+\lambda^M (1-q)}$, $p_1\leq p_2$ implies that such a condition will not be satisfied.

	    \item (existence) If $p_2>\frac{(2q-1)(1-q)\lambda^M
	    L+(q^2+q-1)(1-q)\lambda^M\lambda^V L-\lambda^V (1-q)^2 L}{(\lambda^M)^2[(1-q)^2q]+\lambda^M q (1-q)} $,
	    
	    the set $\left(  \frac{q p_2+\lambda^M (1-q)qp_2+(1-q)L+\lambda^V (1-q)q L}{1+\lambda^M (1-q)}, \frac{(\lambda^V+1)(1-q)L+q(\lambda^M+1)p_2}{1+q\lambda^M} \right)$ is nonempty.
	    
	    On the other hand, if $p_2\leq \frac{(2q-1)(1-q)\lambda^M
	    L+(q^2+q-1)(1-q)\lambda^M\lambda^V L-\lambda^V (1-q)^2 L}{(\lambda^M)^2[(1-q)^2q]+\lambda^M q (1-q)}$, then $p_2<(\lambda^V+1)L$, which implies that \emph{pre-purchasing} is a PE whenever $p_1\leq p_2$.

	    \item  If $p_1\leq (1-q)L+\lambda^V q (1-q)L +qp_2+\lambda^M(1-q)q p_2 $, 
        to pre-purchase, as a PE, gives a higher utility than the PE assigns the consumer to wait.
        
        \item In sum, in the PPE, when $p_1 \leq p_2$ and 
        \begin{align*}
            p_1\leq & \min \bigg\{ \frac{(\lambda^V+1)(1-q)L+q(\lambda^M+1)p_2}{1+q\lambda^M},\\
            &(1-q)L+\lambda^V q (1-q)L +qp_2+\lambda^M(1-q)q p_2 \bigg\}
        \end{align*}
        the consumer will pre-purchase.

	\end{itemize}

    \item $p_2>H$
    
    \begin{itemize}
        \item In the range of $\left(\frac{\mathbb{E}[\omega]}{1+\lambda^M}, \mathbb{E}[\omega]+\lambda^V q^2 H +\lambda^V (1-q^2)L \right)$, both \emph{pre-purchasing} and \emph{waiting} can be supported by some PEs.
        \item If $p_1\leq \mathbb{E}[\omega] -\lambda^V(1-q)q (H-L)$, to pre-purchase, as a PE, gives a higher utility than the PE assigning the consumer to wait. 
        
        \item In sum, in the PPE, the consumer will pre-purchase when
        \begin{equation*}
        \begin{array}{ll}
             p_1 & \leq \min\{ \mathbb{E}[\omega] -\lambda^V(1-q)q (H-L), \mathbb{E}[\omega] + \lambda^V q^2 H + \lambda^V (1-q^2)L\} \\   
             &=\mathbb{E}[\omega] -\lambda^V(1-q)q (H-L).
        \end{array}
        \end{equation*} 
    \end{itemize}
 
 \end{itemize}

\end{proof}

\begin{proof}[Proof of Proposition \ref{prop:mainresult}]
	
    The seller offers $(p_1,p_2)$:
	\begin{itemize}
		\item 
		If $p_1$ is higher than $T_1$-consumer's acceptable price and $p_2\leq L$, the consumer will purchase at $T_2$.

		\item If $p_1$ is higher than $T_1$-consumer's acceptable price and $p_2\in (L,H]$, the consumer will purchase at $T_2$ whenever she realizes that $\omega=H$.

		\item If $p_1$ is higher than $T_1$-consumer's cutoff and $p_2>H$, the consumer will never purchase.

	\end{itemize}
	
	All three of these cases provide the seller an expected profit lower than $\mathbb{E}[\omega]$, so they are not optimal as the following provides a pricing at which the seller earns more than $\mathbb{E}[\omega]$.

According to Lemma \ref{lemma:monotone},
\begin{equation*}
    \max_{p_2}\hat{p}_1(p_2)=\min\{\frac{(\lambda^V+1)(1-q)L+(\lambda^M+1)q\cdot H}{1+q\cdot \lambda^M}, (\lambda^V q+1)(1-q)L+\lambda^M (1-q) q\cdot H+q\cdot H  \}.
\end{equation*}
Because $\hat{p}_1(H)>\mathbb{E}[\omega]$,
\begin{align*}
    &(p^{\ast}_1,p^{\ast}_2)=\\
    &\left( \min\{\frac{(\lambda^V+1)(1-q)L+(\lambda^M+1)q\cdot H}{1+q\cdot \lambda^M},   (\lambda^V q+1)(1-q)L+\lambda^M (1-q) q\cdot H+q\cdot H  \}, H\right)
\end{align*}

is the optimal policy given that the seller commits to the spot price.
	
\end{proof}

\begin{proof}[Proof of Proposition \ref{prop_sec_4.1}]
According to Proposition \ref{prop:standard}, given that the seller makes a spot-price commitment, the seller obtains $\mathbb{E}[\omega]$ by (advance) selling the good to the consumer in the advance-purchase stage. Now suppose the seller chooses not to commit and to adjust the spot price after the seller realizes the state of nature in the spot-market stage. Because the seller will perfectly learn $\omega$, he can then adjust the corresponding price conditional on the state: $p_2^{\omega}=\omega$.
Expecting such pricing, the consumer’s cutoff advance price is $\mathbb{E}[\omega]$. Selling the good at this price guarantees the seller an expected profit at the level of $\mathbb{E}[\omega]$. According to the proof of Proposition \ref{prop:standard}, such a level is the maximum ex-ante profit the seller can achieve when facing a standard consumer.

\end{proof}

\begin{proof}[Proof of Lemma \ref{prop:optpricingnocommit}]

    According to Lemma \ref{lemma:2ndstage}, the $T_2$-consumer's cutoff price equals $\omega$. When optimal, the seller adjusts the spot price to be $p_2^\omega=\omega$.

    \begin{itemize}
        \item[(i)] \emph{Pre-purchase} is supported by a PE if:
    \begin{align*}
        &\mathbb{E}[\omega]-p_1-\lambda^V q (1-q)(H-L) \\ \geq  &\mathbb{E}[\omega]- \mathbb{E}[\omega]-\lambda^V q (1-q)(H-L) -\lambda^M q[H-p_1]^+-\lambda^M(1-q)[L-p_1]^+.
    \end{align*}

    \begin{itemize}
        \item If $p_1>H$, the condition does hold.
        \item If $p_1\leq L$, the condition always holds.
        \item If $p_1\in (L,H]$, then the condition holds if $p_1\leq \frac{\mathbb{E}[\omega]+\lambda^M q H}{1+\lambda^M q }$, where $ \frac{\mathbb{E}[\omega]+\lambda^M q H}{1+\lambda^M q }$ is located within $(\mathbb{E}[\omega],H)$.
        
    \end{itemize}
    
    \item[(ii)] \emph{Wait} is supported by a PE of $T_1$-consumer if:
    \begin{align*}
        &\mathbb{E}[\omega]-p_1-\lambda^V q (1-q)(H-L) -\lambda^M q[p_1-H]^+-\lambda^M (1-q)[p_1-L]^+\\&< \mathbb{E}[\omega]- \mathbb{E}[\omega]-\lambda^V q (1-q)(H-L) -\lambda^M q(1-q)(H-L).
    \end{align*}
  
    \begin{itemize}
        \item If $p_1>H$, the condition always holds.
        \item If $p_1<L$, the condition does not hold.
        \item If $p_1\in (L,H]$, then the condition holds if $p_1>\frac{\mathbb{E}[\omega]+\lambda^M q(1-q)H}{1+\lambda^M q(1-q)}$, where $\frac{\mathbb{E}[\omega]+\lambda^M q(1-q)H}{1+\lambda^M q(1-q)}$ is located within $(\mathbb{E}[\omega],H)$.
    \end{itemize}

    \end{itemize}
    Moreover, $\frac{\mathbb{E}[\omega]+\lambda^M q(1-q)H}{1+\lambda^M q(1-q)}< \frac{\mathbb{E}[\omega]+\lambda^M q H}{1+\lambda^M q }$. Thus, there are multiple PEs.

If $p_1< \mathbb{E}[\omega]+\lambda^M q(1-q)(H-L)$, then pre-purchasing is supported by a PE and is preferred by the $T_1$-consumer among all (if multiple) PEs.

\end{proof}

\begin{proof}[Proof of Proposition \ref{prop: to_commit}]

If $\lambda^V>0$, the following two strict inequalities hold:

\begin{align*}
    \frac{\mathbb{E}[\omega]+q\lambda^M H}{1+q\lambda^M}&<\frac{\mathbb{E}[\omega]+q\lambda^M H+ (1-q)\lambda^V L}{1+q\lambda^M} \\
    \frac{\mathbb{E}[\omega]+q\lambda^M H}{1+q\lambda^M} &<\mathbb{E}[\omega]+\lambda^V q (1-q)L+\lambda^M q (1-q)H
\end{align*}
\end{proof}

\begin{proof}[Proof of Proposition \ref{prop:riskaverse}]
Let $v(k)$ captures a risk-averse consumer's utility given the outcome $k\in \mathbb{R}$, where $v:\mathbb{R}\to \mathbb{R}$ is monotonic increasing and strictly concave.

\noindent Suppose the seller commits to $p_2$. He expects to earn strictly less than $\mathbb{E}[\omega]$: A consumer who has chosen to wait will purchase at $T_2$ whenever $p_2\leq \omega$. Because the loss-averse consumer now encounters no uncertainty at $T_2$, her behavior is similar to that of a risk-neutral consumer. Thus, under optimal pricing, a consumer expects utility $v(0)$ if she chooses to wait at $T_1$. The condition for such a consumer to pre-purchase is
\begin{equation}\label{eq:v-indifferent}
    q\cdot  v((H-p_1))+ (1-q)\cdot v(L-p_1)\geq v(0).
\end{equation}
By the concavity of $v$ and the Jensen's inequality, $q\cdot v((H-\Tilde{p}_1))+ (1-q) \cdot v(L-\Tilde{p}_1)=v(0)$ implies that $\Tilde{p}<\mathbb{E}[\omega]$. The seller will make make this constraint binding, for otherwise, he expects at most $\max\{qH,L\}$, which is less than $\mathbb{E}[\omega]$, by selling the good at $T_2$.

\noindent
Suppose, on the other hand, the seller does not commit. In this case, the seller can adjust the spot price according to $\omega$ at $T_2$. In particular, he can offer a high $p_1$ so that the consumer will wait at $T_1$, then set $p^\omega_2=\omega$ and expects a profit equal to $\mathbb{E}[\omega]$.
By doing so, the seller maximizes his expected profit.

\end{proof}

\begin{proof}[Proof of Proposition \ref{prop:fast_play_pre-purchasing}]

    In light of the example below the proposition. 
    Suppose that the $T_1$-consumer is offered $(p'_1,p'_2)=(\mathbb{E}[\omega], H)$. The expected payoff for the consumer to pre-purchase, given her reference distribution formed by the same plan, is:
\begin{equation*}
\mathbb{E}[\omega]-p'_1 - q\cdot (1-q) \lambda^V (H-L)=- q\cdot (1-q) \lambda^V (H-L).
\end{equation*}

Next, consider a never-purchase plan, $\alpha_c^{np}(\cdot |\bar{h})$ at $\bar{h}:=(p'_1,p'_2)$, so that the consumer chooses to wait at $T_1$ and chooses not to buy at $T_2$ for each $\omega\in \{H,L\}$. Let $\alpha_c^{np}$ represent the first-order belief aligned with this plan.
Then reference distribution is formed by $\alpha_c^{np}(\cdot |\bar{h})$, yielding an expected utility of zero at $\bar{h}$, denoted as $\Tilde{u}{c,\bar{h}}=0$. Moreover, for each $h'=((p'_1,p'_2),H,wait)$ and $h''=((p'_1,p'_2),L,wait)$, the expected utilities are also equal to zero.

Given that $p_2\geq \bar{p}_2(\lambda^V,\lambda^M)=\frac{\mathbb{E}[\omega]-(1-q)\lambda^Vq(H-L)}{1+\lambda^M}$, if $p_1\geq \frac{\mathbb{E}[\omega]-(1-q)\lambda^Vq(H-L)}{1+\lambda^M}$, it can be verified that the never-purchase plan is consistent with a PE. Thus, it is a PE providing a value of $0$ which is strictly greater than $- q\cdot (1-q) \lambda^V (H-L)$. According to Remark \ref{remark:PPE_in_static}, under PPE the $T_1$-consumer will never choose to ``pre-purchase'' at $T_1$ under Assumption \ref{assu:initial_belief}.

The remaining proof involves cases where the never-purchase plan is not validated as a PE, specifically when $p_1<\frac{\mathbb{E}[\omega]-(1-q)\lambda^Vq(H-L)}{1+\lambda^M}$. In such cases, the low $p_1$ will attract the $T_1$-consumer deviates to pre-purchase.
Note, however, that the right-hand side is weakly lower than $\mathbb{E}[\omega]-(1-q)\lambda^Vq(H-L)$, and this inequality becomes strict when $\lambda^M>0$. Consequently, the statement of the proposition is inherently satisfied in such cases.

\end{proof}

\clearpage

\end{document}